\pdfoutput=1
\RequirePackage[T1]{fontenc}
\documentclass[12pt]{article}

\usepackage[height=8.85in,width=6.25in]{geometry}

\usepackage[utf8]{inputenc}
\usepackage{amsmath}
\usepackage{amssymb}
\usepackage{mathtools}
\numberwithin{equation}{section}
\usepackage{slashed}
\usepackage{braket}
\usepackage[svgnames]{xcolor}
\usepackage[colorlinks,citecolor=DarkGreen,linkcolor=FireBrick,urlcolor=FireBrick,linktocpage,breaklinks=true]{hyperref}
\usepackage{cite}
\usepackage{graphicx}
\usepackage{tikz}
\usepackage{times}
\usepackage{courier}
\usepackage{bm}
\usepackage{subfig}
\usepackage{fnpct}
\usepackage[whole]{bxcjkjatype} 

\DeclareMathOperator{\CS}{CS}
\DeclareMathOperator{\PD}{PD}
\DeclareMathOperator{\rank}{rank}
\DeclareMathOperator{\tr}{tr}

\def\CP{\mathbb{CP}}

\def\sB{\mathsf{B}}
\def\sC{\mathsf{C}}
\def\sG{\mathsf{G}}
\def\bC{\mathbb{C}}
\def\bR{\mathbb{R}}
\def\bZ{\mathbb{Z}}
\def\Nequals#1{$\mathcal{N}{=}#1$}
\def\u#1{\underline{#1}}
\def\cD{\mathcal{D}}
\def\cT{\mathcal{T}}
\def\cO{\mathcal{O}}
\def\cI{\mathcal{I}}

\def\cA{\mathcal{A}}
\def\cB{\mathcal{B}}
\def\cH{\mathcal{H}}
\def\One{\mathbf{I}}
\def\tildetimes{\mathrel{\tilde\times}}

\begin{document}

\begin{titlepage}

\begin{flushright}
\end{flushright}

\vskip 3cm

\begin{center}
\renewcommand*{\thefootnote}{\fnsymbol{footnote}}
{\Large\bfseries On holography with ADE singularities}

\vskip 1cm
Sunjin Choi\footnotemark[2]\footnotetext[2]{\ sunjin.choi@ipmu.jp} and
Yuji Tachikawa\footnotemark[8]\footnotetext[8]{\ yuji.tachikawa@ipmu.jp}
\vskip 1cm

\begin{tabular}{ll}
 & Kavli Institute for the Physics and Mathematics of the Universe (WPI), \\
& University of Tokyo,  Kashiwa, Chiba 277-8583, Japan
\end{tabular}

\vskip 2cm

\end{center}

\noindent 

We study aspects of the AdS/CFT correspondence 
for \Nequals4 $U(N)$ super Yang-Mills theory on  $S^3/\Gamma$,
where $\Gamma \subset SU(2)$ is a finite subgroup
leading to an ADE singularity in the bulk AdS geometry.
We show that
a large vacuum degeneracy arises from the choice of gauge holonomy on $S^3/\Gamma$.
On the gravity side, we argue that the bulk ADE singularity supports topological degrees of freedom
responsible for this degeneracy.
We then provide a holographic derivation of a corresponding large vacuum degeneracy 
for class S theories of type $U(N)$,
showing that these topological degrees of freedom admit an effective description
in terms of a three-dimensional level-$N$ Chern-Simons theory,
whose gauge group $\sG$ is determined by $\Gamma$.
Finally, we discuss how the one-form symmetries of the \Nequals4 super Yang-Mills theory
are realized on the Chern-Simons theory side.

\end{titlepage}

\setcounter{tocdepth}{2}
\tableofcontents

\section{Introduction and summary}
\label{sec:intro}

\subsection{Overview}
The AdS/CFT correspondence
has helped us learn 
various interesting and intricate features of both quantum field theory and quantum gravity
from its inception \cite{Maldacena:1997re},
even in its very prototypical form 
of the duality between \Nequals4 $U(N)$ super Yang-Mills theory
and Type IIB string theory on AdS$_5\times S^5$.
There are diverse extensions of this basic duality in a multitude of directions,
but the analysis of the effects of spacetime topology on the field theory side 
has attracted relatively little attention in the authors' opinion. 
This paper is intended to fill some of these gaps in the literature.

We start by considering \Nequals4 $U(N)$ super Yang-Mills theory on $S^3/\Gamma$.
To preserve a half of the supersymmetries,
we take $\Gamma$ to be a finite subgroup of $SU(2)\simeq S^3$,
which are famously classified by ADE types.
This system is known  \cite{Horowitz:2001uh,Lin:2005nh} to have a large number of very light states
labeled by 
flat bundles on $S^3/\Gamma$.
It was further noted later that, to a varying degree of generality,
these light states are actually exactly degenerate \cite{Razamat:2013opa,Ju:2023ssy}.
A string duality chain  to explain it was given in \cite{Ju:2023umb}, by mapping
this degeneracy to that of the ground states on $T^2$ of three-dimensional level-$N$ pure Chern-Simons theory 
whose gauge group $\sG$ is determined by $\Gamma$
by the McKay correspondence.

Holographically, we have Type IIB string theory on $\bR_t\times B^4/\Gamma \times S^5$
with $N$ units of flux through $S^5$,
where $\bR_t$ is the time direction and $B^4/\Gamma$ is a quotient of 
the four-dimensional ball $B^4$.
Interestingly, there is an ADE singularity at the origin of $B^4/\Gamma$,
although the boundary $S^3/\Gamma$ is smooth.
This means that the ADE singularity in the bulk provides a large degeneracy in holography.

In this paper, we first firmly establish this exact degeneracy of the holonomy sectors of
\Nequals4 super Yang-Mills theory on $S^3/\Gamma$ on the field theory side,
by providing a uniform  derivation of the vanishing of the supersymmetric Casimir energy
applicable to all finite subgroups $\Gamma$ of $SU(2)$ and to arbitrary gauge groups.
Unfortunately, the origin of the large degeneracy on the bulk holographic side
is still somewhat mysterious.

A clue is provided by a large degeneracy of a similar nature 
in the case of M-theory on $\bR_t\times B^4/\Gamma \times \Sigma \tildetimes S^4$
with $N$ units of flux through $S^4$,
where $\Sigma$ is a Riemann surface of genus $g>1$ and we mean by the notation 
$\tildetimes S^4$  that $S^4$ is nontrivially fibered over $\Sigma$.
This geometry, originally constructed in \cite{Maldacena:2000mw} and further quotiented by $\Gamma$,
is holographically dual to the four-dimensional \Nequals2 class S theory of type $U(N)$, arising from the six-dimensional \Nequals{(2,0)} theory of type $U(N)$ compactified on $\Sigma$, on $S^3/\Gamma$.
In this case, it is straightforward to relate this degeneracy on the holographic side to
the vacuum degeneracy of three-dimensional pure Chern-Simons theory at level $N$,
whose gauge group $\sG$ is determined by $\Gamma$ via its ADE classification.

Field-theoretically, \Nequals4 super Yang-Mills theory is a class S theory for $\Sigma=T^2$.
Although the M-theory holographic background of \cite{Maldacena:2000mw} is not applicable when $\Sigma=T^2$, 
it seems reasonable to assume that the vacuum degeneracy on $S^3/\Gamma$ of \Nequals4 super Yang-Mills theory is similarly explained by that of 3d pure Chern-Simons theory on $T^2$
with gauge group $\sG$.\footnote{%
In \cite{Lin:2005nh},  \Nequals4 $U(N)$ super Yang-Mills theory on $S^3/\bZ_k$
was already related to 3d $U(N)_k$ Chern-Simons theory by T-duality
along the Hopf fiber of $S^3/\bZ_k$.
A further relation to its level-rank dual, which is the pure $U(k)_N$ Chern-Simons theory,
was also briefly discussed there.
The analysis in \cite{Lin:2005nh} did not carefully treat the $U(1)$ parts of these theories, however,
as the authors of \cite{Lin:2005nh} themselves mentioned there.
In this paper, we 
treat the issue of the $U(1)$ part and the supersymmetric Casimir energy more carefully.
Also, we  generalize the construction from Abelian subgroups $\bZ_k\subset SU(2)$ to
more general finite subgroups $\Gamma\subset SU(2)$.
}\footnote{
We should also mention that the duality between 
\Nequals4 super Yang-Mills and more general class S theories on $S^3/\Gamma$
and 3d pure Chern-Simons theory with gauge group $\sG$ was already provided
in \cite{Ju:2023umb,Ju:2023ssy,Albrychiewicz:2024fkr},
by directly analyzing D3-branes or M5-branes on $\bC^2/\Gamma$ singularity.
In this paper,  the analysis will be done on the holographic dual,
as this allows us to avoid dealing with branes hitting the singularity,
which might provide unwanted additional degrees of freedom.
That said, the technical part of the duality does not differ 
in any significant manner from \cite{Ju:2023umb,Ju:2023ssy,Albrychiewicz:2024fkr},
and the authors of this paper do not claim any originality on this point.
\label{foot:1st}
} 

In the analysis that follows, we will carefully examine how the 1-form symmetries of \Nequals4 super Yang-Mills theory and more general class S theories are mapped to 
the 1-form symmetry of 3d pure Chern-Simons theory.
In particular, we provide a detailed check of their actions in the case of \Nequals4 super Yang-Mills theory on $S^3/\Gamma$ and 3d Chern-Simons theory on $T^2$.

\subsection{Detailed introduction}
With this short overview given, let us provide more details. 
\paragraph{Historical context.} 
Although many studied the  AdS/CFT duality on AdS$_5\times X_5$, where $X_5$ is more general than $S^5$,
immediately after \cite{Maldacena:1997re},
the first to tweak the side of AdS$_5$ to the authors' knowledge 
was \cite{Horowitz:2001uh}. 
There, it was noticed that there are very light additional states 
when the boundary field theory is on any space of the form $M/\Gamma$, where $\Gamma$ acts without fixed points on $M$.
This is due to flat connections, whose holonomies are specified by homomorphisms, 
\begin{equation}
\rho\colon \Gamma\to U(N)\ .\label{intro-foo}
\end{equation}
The number of such states therefore scales as $O(N^{\ell_\Gamma-1})$,
where $\ell_\Gamma$ is the number of irreducible representations of $\Gamma$.
The case of $S^3/\bZ_k$ was then studied in detail in \cite{Lin:2005nh},
e.g.~by constructing smooth holographic dual geometries for each of the representations $\rho:\bZ_k\to U(N)$.

One subtlety not addressed in these early papers was the supersymmetric Casimir energy
due to the zero-point energies of various fields in the system.
As the holonomy affects the boundary conditions satisfied by the fields,
the zero-point energies can depend on it,
and the cancellation between bosonic and fermionic contributions is not necessarily perfect
even in the presence of supersymmetry when the background is curved.

For $S^3/\bZ_k$, this holonomy-dependent supersymmetric Casimir energy was computed in \cite{Benini:2011nc} for general
\Nequals1 supersymmetric theories, where it was shown that they are generically nonzero.
This later played an important role in the check of Seiberg duality on $S^3/\bZ_k$ in \cite{Razamat:2013opa}.
There, the Casimir energies of \Nequals4 $U(N)$ super Yang-Mills theories  on $S^3/\bZ_N$
for $N=3,4,6$ were found to be independent of the holonomy by an explicit computation.
This analysis was extended to \Nequals4 super Yang-Mills theory with various other gauge groups on
various other quotients $S^3/\Gamma$ in \cite{Ju:2023ssy}, where the Casimir energy was found to be
zero whenever it was computed. 
In particular, this analysis established that $O(N^{k-1})$ states labeled by the holonomy studied by \cite{Lin:2005nh} were actually perfectly degenerate.
Our first result in this paper is a uniform derivation of this exact degeneracy on the field theory side
for \Nequals4 super Yang-Mills theory on $S^3/\Gamma$,
valid for all gauge groups $G$ and all finite subgroups $\Gamma\subset SU(2)$.

A chain of string dualities was described in \cite{Ju:2023umb}
to explain this perfect degeneracy,
by mapping it to the degeneracy of three-dimensional pure Chern-Simons theory with gauge group $\sG$ of level $N$ on $T^2$,
where $\sG$ is determined from $\Gamma$ using the McKay correspondence, see Table~\ref{tab:McKay}.
This duality chain was then extended to the case of more general class S theories 
for a Riemann surface $\Sigma$ of genus $g>1$ in \cite{Albrychiewicz:2024fkr},
where some field theoretical checks of the resulting degeneracy was also provided.
Our second aim in this paper is to give a holographic version of the same analysis,
while paying careful attention to how the 1-form symmetries on the four-dimensional side
will be mapped to the 1-form symmetry of the pure Chern-Simons theory.

\begin{table}
\[
\begin{array}{c||c|c|c|c|cccccc}
\Gamma &  \bZ_k & \hat \cD_k &  \hat \cT & \hat \cO & \hat \cI\\
\hline
\sG & A_{k-1}= SU(k) & D_{k+2}=Spin(2k+4) & E_6 & E_7 & E_8\\
\hline
\ell _\Gamma-1= r_\sG & k-1 & k+2 & 6 & 7 & 8 \\
\hline 
I_\Gamma=Z_\sG & \bZ_k & \bZ_4 , \ \bZ_2\times \bZ_2 & \bZ_3 & \bZ_2 & \bZ_1
\end{array}
\]
\caption{The relation between the finite subgroup $\Gamma$ of $SU(2)$
and the simply connected compact gauge group $\sG$ of the dual Chern-Simons theory. 
Here, $\hat \cD_k$ is the binary dihedral group with $4k$ elements,
and $\hat \cT$, $\hat \cO$ and $\hat \cI$ are the binary tetrahedral, octahedral and icosahedral groups.
$\ell_\Gamma$  is the number of irreducible representations of $\Gamma$,
and $r_\sG$ is the rank of $\sG$. They satisfy $\ell_\Gamma-1=r_\sG$.
Finally, 
$I_\Gamma=Z_\sG$ is the group of one-dimensional representations of $\Gamma$,
or the center of $\sG$.
For $\hat \cD_k$ or equivalently for $Spin(2k+4)$, $I_\Gamma=Z_\sG$ is $\bZ_4$ or $\bZ_2\times \bZ_2$ depending on whether $k$ is odd or even.
\label{tab:McKay}}
\end{table}

\paragraph{Duality to three-dimensional Chern-Simons theory.}
Let us now outline the duality, which is a modified version of what was presented in \cite{Ju:2023ssy} to better match our purposes.
Our gauge theory lives on $\bR_t \times S^3/\Gamma$,
where $\bR_t$ is the time direction.
For \Nequals4 super Yang-Mills theory,
the holographic dual is Type IIB string theory on $\bR_t \times B^4/\Gamma \times S^5$, where $B^4$ is a four-dimensional ball.
One distinguishing feature is that $B^4/\Gamma$ has an orbifold singularity at the origin of the form $\bC^2/\Gamma$,
even though the boundary spacetime $\bR_t\times S^3/\Gamma$ is smooth.
The natural guess then is that the degeneracy arises from the localized degrees of freedom on this singular locus,
since the infrared limit on the boundary theory should be holographically mapped to the region deep inside the bulk.

It is well-known that the six-dimensional \Nequals{(2,0)} theory of type $\Gamma$ lives on the 
$\bC^2/\Gamma$ singularity of Type IIB string theory.
Then, the holographic dual has this six-dimensional theory wrapped on $S^5$, further coupled to 
the RR four-form field with flux $\int_{S^5} F_5=N$.
As the coupling of the \Nequals{(2,0)} theory to the background five-form flux has not been studied to the authors' knowledge,
this description does not help.
We can still see that we are on the right track, by the following consideration.

If we blow up the singularity slightly, it is known that there are small two-dimensional spheres 
intersecting according to the Dynkin diagram of $\sG$.
On the five-dimensional bulk, there is a topological degree of freedom described by the action \begin{equation}
N \int_{M_5} B_2 dC_2\ ,
\end{equation} as was first discussed in \cite{Witten:1998wy},
which is now often discussed in the context of SymTFT describing the $U(1)^2$ one-form symmetry of \Nequals4 $U(N)$ super Yang-Mills theory, see e.g.~\cite{Bergman:2022otk}.
Let us denote the small 2-spheres by $S^2_{i}$ with $i=1,\ldots,r_\sG$,
where $r_\sG$ is the rank of $\sG$.
Then $p_i:=\int_{S^2_i} B_2$ and $q_i:=\int_{S^2_j} C_2$ provide canonically conjugate pairs of 
localized degrees of freedom, with the action $~N\int C^{ij} p_i \dot q_j dt$,
where $C^{ij}$ is the Cartan matrix.
Applying Bohr-Sommerfeld quantization, we see that there are $\sim N^{r_\sG}$ states,
correctly reproducing the  form of the degeneracy
arising from the choice of holonomies given in \eqref{intro-foo}.

Still, it is unclear how to determine the exact number of the degenerate states.
Instead, we consider the vacuum degeneracy of a related system,
namely class S theories of type $U(N)$,
obtained by wrapping $N$ M5-branes, with the center-of-mass modes included,
on a Riemann surface $\Sigma$ of genus $g$
and taking the infrared limit. 
In this case, the holographic dual was constructed by Maldacena and Nu\~nez in \cite{Maldacena:2000mw}, and has the form of $M_5\times \Sigma \tildetimes S^4$
with $N$ units of flux through $S^4$.
Here, we mean by the notation $\tildetimes S^4$
that $S^4$ is fibered nontrivially over $\Sigma$,
and we will take $M_5=\bR_t \times B^4/\Gamma$.

M-theory on $B^4/\Gamma$ is known to have seven-dimensional gauge fields with gauge group $\sG$ on the singularity,
with the coupling \begin{equation}
\sim \int_{7d} G_4 \CS_\sG (A)\ ,
\end{equation} where $G_4$ is the M-theory four-form flux and $\CS_\sG (A)$ is the Chern-Simons form for $\sG$.
Using $\int_{S^4} G_4=N$, we now have the level-$N$ pure Chern-Simons theory with gauge group $\sG$ 
on $\bR_t\times \Sigma$.

Note that \Nequals4 super Yang-Mills theory is, field-theoretically speaking,
the class S theory when $\Sigma=T^2$. 
Although the Maldacena-Nu\~nez background is not applicable in this case,
it seems reasonable that the same three-dimensional Chern-Simons theory, 
but now with $\Sigma=T^2$,
governs the vacuum degeneracy of \Nequals4 $U(N)$ super Yang-Mills theory on $S^3/\Gamma$.
Indeed, the states of this Chern-Simons theory on $T^2$ is labeled by 
irreducible unitary integrable representations of affine Lie algebra of type $\sG$ at level $k$,
which is known to be in one-to-one correspondence with \eqref{intro-foo},
as we will describe in more detail in Sec.~\ref{sec:states}.

\paragraph{One-form symmetry.}

In this paper, we will also study how the 1-form symmetries of \Nequals4 super Yang-Mills theory and more general class S theories on $S^3/\Gamma$
are mapped to the 1-form symmetry of the pure $\sG$ Chern-Simons theory on $T^2$ or more general surfaces $\Sigma$.
Concentrating to the case of \Nequals4 super Yang-Mills theory here,
recall first that \Nequals4 $U(N)$ super Yang-Mills theory has two 1-form symmetries,
one electric and one magnetic \cite{Gaiotto:2014kfa}.
The Hilbert space of the theory on $S^3/\Gamma$ then has a (generally projective) action of $I_\Gamma\times I_\Gamma$,
where $I_\Gamma$ is the finite Abelian group of one-dimensional representations of $\Gamma$.
Now, the dual pure $\sG$ Chern-Simons theory also has a 1-form symmetry, 
which is given by the center $Z_\sG$ of $\sG$.
This results in the (generally projective) action of $Z_\sG\times Z_\sG$ on the Hilbert space of states on $T^2$.
We will check explicitly that $I_\Gamma=Z_\sG$, and the actions of these groups across the duality are consistent.

Note also that two factors of $I_\Gamma\times I_\Gamma$ are exchanged 
by the S-duality of \Nequals4 super Yang-Mills theory,
while two factors of $Z_\sG\times Z_\sG$ are exchanged 
by the geometric S-transformation on $T^2$.
We will see that these two actions on the states of the system are again compatible.

In our analysis, the 1-form symmetry \emph{groups} on both sides of duality are different:
we have $U(1)^2$ on the side of \Nequals4 super Yang-Mills theory,
and $Z_\sG$ on the side of the pure Chern-Simons theory.
Still, the actions on the Hilbert spaces are consistent, by further taking into account 
that the dual theories are formulated on different manifolds, namely $S^3/\Gamma$ and $T^2$, respectively.
The identification of the symmetry generators will be made possible by
following how the duality maps the background fields of the 1-form symmetry 
on one side of the duality to the other.

\subsection{Organization of the paper}

The rest of the paper is organized as follows. \begin{itemize}
\item
In Sec.~\ref{sec:casimir}, we compute the supersymmetric Casimir energy of
each holonomy sector of four-dimensional \Nequals4 super Yang-Mills theory with gauge group $G$
on $S^3/\Gamma$, where $\Gamma$ is a finite subgroup of $SU(2)$.
We show that it uniformly vanishes, independent of $G$, $\Gamma$ and the chosen holonomy sector.
We will see that the structure of \Nequals4 supersymmetry multiplet is essential in this derivation,
so that the supersymmetric Casimir energy is generically nonzero for less supersymmetric field theories.

\item
In Sec.~\ref{sec:duality}, we interpret this degeneracy of \Nequals4 super Yang-Mills theory on $S^3/\Gamma$ holographically.
We mainly study a related system of the class S theories of type $U(N)$
defined using a Riemann surface $\Sigma$ of genus $g>1$, placed on $S^3/\Gamma$.
We discuss how the ADE singularity at the origin of the holographic dual
eventually leads to 
three-dimensional pure Chern-Simons theory with gauge group $\sG$ and level $N$ on $\Sigma$.
We will pay particular attention to how the background fields for the one-form symmetry
on one side are mapped to those on the other side.

\item
In Sec.~\ref{sec:matching}, we will apply the duality established for Riemann surfaces $\Sigma$
of genus $g>1$  to the case of $\Sigma=T^2$, to study the correspondence
between the holonomy sectors of four-dimensional \Nequals4 $U(N)$ super Yang-Mills theory on $S^3/\Gamma$
and the states of three-dimensional pure Chern-Simons theory on $T^2$
with gauge group $\sG$ at level $N$.
We will see the matching of not only the numbers of states
but also the actions of the one-form symmetries across the duality.

\item The paper concludes with a short discussion of future directions in Sec.~\ref{sec:conclusions}.
\item 
We also have two appendices. 
In  Appendix~\ref{app:geom},  we collect standard facts on the topology
of $S^3/\Gamma$ and the associated asymptotically locally Euclidean (ALE) spaces.
In Appendix~\ref{sec:checks}, we give the details of the McKay correspondence
concerning $I_\Gamma$ and $Z_\sG$, which become necessary in our analysis in Sec.~\ref{sec:1-form-matching}.
\end{itemize}

\section{Supersymmetric Casimir energy}
\label{sec:casimir}
In this section, we study the supersymmetric Casimir energy
in the computation of the supersymmetric index of four-dimensional \Nequals4 super Yang-Mills theory
with arbitrary gauge group $G$ on  $S^3/\Gamma$,
where $\Gamma$ is a finite subgroup of $SU(2)$.
We will find that the supersymmetric Casimir energy is independent of 
the choice of the gauge holonomy $\rho_G: \Gamma\to G$,
for \emph{any} $G$ and $\Gamma$.
Thus, we obtain a vacuum degeneracy parametrized by $\rho_G$.
Note that in this section $G$ stands for the gauge group of four-dimensional super Yang-Mills theory,
and \emph{not} the gauge group of three-dimensional Chern-Simons theory discussed later,
for which we reserve a different symbol $\sG$.

This degeneracy was already noted in \cite{Razamat:2013opa}
when $\Gamma=\bZ_k$ and $G$ has the Lie algebra $su(N)$, for various choices of $(k,N)$ at least.
It was also studied in a case-by-case basis for some $\Gamma$ and $G$ 
in two papers by C. Ju  \cite{Ju:2023umb,Ju:2023ssy}.
The aim here is to give a uniform derivation applicable to all cases.

\subsection{\Nequals4 super Yang-Mills theory on $S^3$}

Let us first start with some general properties of the radially quantized \Nequals4 super Yang-Mills theory on  $\mathbb{R}_t \times S^3$. The full superconformal group is $PSU(2,2|4)$ whose bosonic subgroup is given by $SU(2,2) \times SU(4)_R \cong SO(4,2) \times SO(6)_R$.\footnote{%
In this paper we do not distinguish two connected Lie groups sharing the same Lie algebra,
except when we discuss gauge groups.
}
Its compact subgroup contains $SO(4) \cong SU(2)_l \times SU(2)_r$ rotation symmetry on $S^3$ and $SU(4)_R \cong SO(6)_R$ R-symmetry. In addition, there are 32 fermionic generators: 16 Poincar\'e supercharges $Q^i_\alpha$, $\bar{Q}_{i\dot\alpha}$, and 16 conformal supercharges $S_i^\alpha = (Q^i_\alpha)^\dagger$, $\bar{S}^{i\dot\alpha} = (\bar{Q}_{i\dot\alpha})^\dagger$, which are Hermitian conjugate to each other in the radially quantized setup. Here, $i=1,2,3,4$ is the index for $SU(4)_R$ R-symmetry in the fundamental/anti-fundamental representation, and $\alpha, \dot\alpha$ are the doublet indices for $SU(2)_l, SU(2)_r$ rotation symmetries.

We shall choose one supercharge $Q\equiv Q^4_-$ and focus on the $\frac{1}{16}$-BPS states on $S^3$ annihilated by both $Q$ and $Q^\dagger=S\equiv S_4^-$. They can be organized into the simultaneous eigenstates of energy $E$, $SU(2)_l$ angular momentum $J_l$, and $U(1)_R$ superconformal R-charge $R$, whose eigenvalues satisfy the following BPS relation:
\begin{equation}
    2\,\{Q^\dagger, Q\} = E- \frac{3}{2} R - 2J_l = 0\ .
\end{equation}
Note that in terms of the Cartan charges $R_I$ of $SO(6)_R$ R-symmetry, we have $R \equiv \frac{2}{3}(R_1+R_2+R_3)$. 
The subgroup commuting with $Q$ and $Q^\dagger$ is $PSU(1,2|3) \subset PSU(2,2|4)$ whose bosonic subgroup contains $SU(2)_r$ and $SU(3) \subset SU(3) \times U(1) \subset SU(4)_R$. In the free limit, one can easily construct $\frac{1}{16}$-BPS states satisfying the above relation using gauge-covariant BPS elementary fields, which are all in the adjoint representation of $G$, satisfying the same relation. In the $\mathcal{N}{=}1$ language, they are given by
\begin{equation}
    \bar\phi^m\ , \quad \psi_{m+}\ , \quad f_{++}\ , \quad \bar\lambda_{\dot\alpha}\ ,
\end{equation}
where $m=1,2,3$ is the $SU(3) \subset SU(4)_R$ index either for the fundamental/anti-fundamental representation and we omitted all gauge indices for the adjoint representation. Here, $f_{++}$ is the anti-self-dual component of the field strength. In addition, we can act an arbitrary number of BPS derivatives $\partial_{+\dot\alpha}$ on them. Quantizing these fields, we obtain single-letter (or single-particle) BPS modes on $S^3$ in the free limit. As derivatives $\partial_{+\dot\alpha}$ acting on the same field commute in the free limit, all $SU(2)_r$ indices of $\partial_{+\dot\alpha}$ acting on a single mode should be symmetrized. 
Then, we have the following independent single-letter BPS modes:
\begin{equation}\label{single-mode}
    \partial_{+(\dot\alpha_1|} \cdots \partial_{+|\dot\alpha_n)} \bar\phi^m\, , \;\, \partial_{+(\dot\alpha_1|} \cdots \partial_{+|\dot\alpha_n)} \psi_{m+}\, , \;\, \partial_{+(\dot\alpha_1|} \cdots \partial_{+|\dot\alpha_n)} f_{++}\, , \;\, \partial_{+(\dot\alpha_1|} \cdots \partial_{+|\dot\alpha_n)} \bar\lambda_{\dot\beta}\ .
\end{equation}
Further note that they are constrained by the BPS equation of motion (in the free limit) as
\begin{equation}\label{eom}
    \partial_{+\dot\alpha} \bar\lambda^{\dot\alpha} = 0\ .
\end{equation}
Multiplying \eqref{single-mode} subject to \eqref{eom} and contracting all gauge indices, one obtains general gauge-invariant $\frac{1}{16}$-BPS states on $S^3$ in the free limit. 
For the details, e.g. notations and charges of the elementary fields, etc., we refer the reader to \cite{Kinney:2005ej,Grant:2008sk} and references therein.

We can count the number of single-letter BPS modes \eqref{single-mode} subject to \eqref{eom} in the free limit with various weights as follows \cite{Romelsberger:2005eg,Kinney:2005ej}:
\begin{align}
\tr_{\textrm{single}(S^3)} \left[(-1)^F t^{3(R+2J_l)} y^{2J_r}  g\right] &= \frac{3t^2-(y+y^{-1})t^3-3t^4+2t^6}{(1-yt^3)(1-y^{-1}t^3)} \tr_{\textrm{Ad}_G}g \nonumber \\
&=\left(1-\frac{(1-t^2)^3}{(1-yt^3)(1-y^{-1}t^3)}\right) \tr_{\textrm{Ad}_G}g\ ,
\label{single-ind}
\end{align}
where the operator inside the trace is chosen to anti-commute with our chosen supercharges $Q$ and $Q^\dagger$ so that the contributions from non-BPS modes are pairwise canceled. This is called the single-letter index.\footnote{Its plethystic exponential gives the multi-letter (or multi-particle) index. The Haar measure integral over $g$ then projects onto gauge-invariant states. This yields the superconformal index \cite{Romelsberger:2005eg,Kinney:2005ej} counting general gauge-invariant $\frac{1}{16}$-BPS states on $S^3$. It is invariant under any continuous deformations, including the Yang-Mills coupling.} 
Here, the trace is taken over \eqref{single-mode} subject to \eqref{eom}, $F$ is the fermion number operator, $J_r$ is the angular momentum for $SU(2)_r$ rotation, and $g\in G$. 

Let us explain \eqref{single-ind} in detail. The overall factor $\tr_{\textrm{Ad}_G} g$ appears since all modes are in the adjoint representation of $G$. In the numerator, $+3t^2$ comes from $\bar\phi^m$, $-(y+y^{-1})t^3$ from $\bar\lambda_{\dot\alpha}$, $-3t^4$ from $\psi_{m+}$, $+t^6$ from $f_{++}$, and finally another $+t^6$ is added to compensate the null mode \eqref{eom}, which should not be counted. The denominator comes from the symmetrized derivatives $\partial_{+\dot\alpha}$ acting on these modes as in \eqref{single-mode}. 

\subsection{Supersymmetric Casimir energy on $S^3/\Gamma$}

Now, we consider $S^3/\Gamma$, where $\Gamma$ freely acts on $S^3$, i.e. $S^3/\Gamma$ is smooth.
Here, $\Gamma$ is a finite subgroup of $SU(2)$ in $SO(4)$ rotation as
\begin{equation}
    \Gamma \subset SU(2)_r \subset SU(2)_l \times SU(2)_r \cong SO(4)\ .
\end{equation}
We will focus on $\frac{1}{16}$-BPS states on $S^3/\Gamma$, that is to say, $\frac{1}{16}$-BPS states on $S^3$ that are invariant under the action of $\Gamma$. This $\Gamma$ also breaks half of supersymmetries $\bar Q_{i\dot\alpha},\bar S^{i\dot\alpha}$, thus even the vacuum states on $S^3/\Gamma$ are $\frac{1}{2}$-BPS. 
The subgroup commuting with $Q$, $Q^\dagger$ and $\Gamma$ is $SU(1|3) \subset PSU(1,2|3) \subset PSU(2,2|4)$. Let us consider the supersymmetric index, which receives contributions from $\frac{1}{16}$-BPS states on $S^3/\Gamma$ defined as
\begin{equation}\label{index-gamma}
    Z (t)=\textrm{Tr}_{\mathcal{H}(S^3/\Gamma)} \left[(-1)^F t^{3(R+2J_l)} \right]\ ,
\end{equation}
where now the trace is taken over the full Hilbert space on $S^3/\Gamma$. 
Compared to \eqref{single-ind}, we cannot turn on the fugacity $y$ since $J_r$ does not commute with $\Gamma$ in general, unless $\Gamma = \mathbb{Z}_k$.
While performing the path integral on $S^3/\Gamma \times S^1$ to compute the above index $Z$, one should also turn on the discrete gauge holonomy $\rho_G:\Gamma=\pi_1(S^3/\Gamma) \to G$.
The full index $Z$ is then obtained by summing over the contributions from all sectors labeled by these distinct gauge holonomy assignments (up to conjugacy classes) as $Z(t) = \sum_{\rho_G} Z(\rho_G;t)$.

The path integral at each holonomy sector to compute $Z(\rho_G;t)$ naturally includes the corresponding zero-point energy or supersymmetric Casimir energy \cite{Aharony:2003sx,Kim:2009wb,Imamura:2011su,Benini:2011nc,Kim:2012ava,Kim:2013nva,Martelli:2015kuk}, which is given by
\begin{equation}\label{Casimir}
    \begin{aligned}
    \epsilon_0 (\rho_G)  := 
    \frac{3}{2}\tr_{\textrm{single}(S^3/\Gamma;\rho_G)} \left[(-1)^F (R+2J_l) \right]\ ,
    \end{aligned}
\end{equation}
where 
the trace is taken over the single-letter BPS modes on $S^3/\Gamma$ with the holonomy $\rho_G$  turned on. The charge combination $R+2J_l$ descends from the definition of $Z$ \eqref{index-gamma}, which vanishes for our chosen supercharges $Q$ and $Q^\dagger$, ensuring pairwise cancellation of the contributions from non-BPS modes. This quantity is divergent just as the standard Casimir energy, so it requires appropriate supersymmetry preserving regularization, e.g. via the zeta function regularization. We shall focus on the differences of the supersymmetric Casimir energy between the holonomy sectors, i.e. on the dependence of $E_0$ on $\rho_G$. The overall factor depends on the renormalization scheme through local counterterms \cite{Birrell:1982ix,Assel:2015nca,Panopoulos:2023cdg}, but it will not affect our discussion.

One convenient way to regularize \eqref{Casimir} is to consider  
\begin{equation}\label{single-gamma}
    f(\rho_G;t) = \tr_{\textrm{single}(S^3/\Gamma;\rho_G)} \left[(-1)^F t^{3(R+2J_l)} \right]\ ,
\end{equation}
and take
\begin{equation}
    \epsilon_0 (\rho_G) = -\frac{1}{2} \lim_{\beta \to 0^+} \frac{d}{d\beta} f(\rho_G;t=e^{-\beta})\ ,
\end{equation}
which can be understood as the heat kernel regularization \cite{Martelli:2015kuk}. Here, $\beta$ is a regulator parameter that will be taken to zero at the end. In general, we need to disregard divergent terms in $\beta$ if any, which should be canceled by appropriate counterterms  \cite{Aharony:2003sx,Kim:2013nva}. 
In our \Nequals4 case, we will see that there are no divergent terms to start with, thanks to high degree of supersymmetry.

The single-letter index $f(\rho_G;t)$ on $S^3 /\Gamma \times S^1$ with the discrete gauge holonomy $\rho_G:\Gamma \to G$ receives contributions only from the $\Gamma$-invariant modes of \eqref{single-mode},
where $\Gamma$ acts both geometrically via $\Gamma\subset SU(2)_r$ and also via the gauge holonomy $\rho_G$. Projecting \eqref{single-mode} onto the singlet of $\Gamma$ applying the projector:
\begin{equation}
P_\textrm{singlet} = \frac{1}{|\Gamma|} \sum_{g\in \Gamma} \rho_\textrm{full} (g)\ ,
\end{equation}
where $\rho_\text{full}$ includes both the geometric and gauge action, and taking the trace over it as \eqref{single-gamma}, we obtain
\begin{equation}
f(\rho_G;t) 
=\frac{1}{|\Gamma|} \sum_{g\in \Gamma} \tr_{\textrm{single}(S^3)} \left[(-1)^F t^{3(R+2J_l)} \rho_{\bC^2}(g)  \rho_G(g)\right]\ .
\end{equation}
Here, $\rho_{\bC^2}(g)$ is the two-dimensional representation of $\Gamma\subset SU(2)_r$ acting on $\bC^2$.
Inside the trace, we can diagonalize $\rho_{\bC^2}(g)$ as \begin{equation}
\rho_V(g) \sim \begin{pmatrix}
\lambda(g) & 0\\
0& \lambda(g)^{-1}
\end{pmatrix}\ ,
\end{equation}
meaning that the factor $y^{2J_r}$ 
in \eqref{single-ind} should be replaced by  $\lambda(g)^{\pm1}$,
so we have 
\begin{equation}
\begin{aligned}
f(\rho_G;t) 
&=\frac{1}{|\Gamma|} \sum_{g\in \Gamma} \left(1-\frac{(1-t^2)^3}{(1-\lambda(g)t^3)(1-{\lambda(g)^{-1}}t^3)} \right) 
\tr_{\textrm{Ad}_G} \rho_G(g)\ .
\end{aligned}
\end{equation}

Now, for any $g$ with $\lambda (g) \neq 1$, we find
\begin{equation}
    1-\frac{(1-t^2)^3}{(1-\lambda(g)t^3)(1-{\lambda(g)^{-1}}t^3)}  = 1+ O(\beta^3)\ ,
\end{equation}
in the limit where $\beta \to 0$, 
yielding only vanishing contributions to the supersymmetric Casimir energy. On the other hand, when $\lambda(g)={\lambda(g)^{-1}}=1$, i.e.~when $g$ is the identity element, we find instead
\begin{equation}
    1-\frac{(1-t^2)^3}{(1-\lambda(g)t^3)(1-{\lambda(g)^{-1}}t^3)} = 1- \frac{(1-t^2)^3}{(1-t^3)^2} = 1+ O(\beta)\ ,
\end{equation}
in $\beta \to 0$ limit.
Therefore, only the identity element of $\Gamma$ contributes to the supersymmetric Casimir energy and we obtain
\begin{equation}
\epsilon_0  = -\frac{1}{2} \lim_{\beta \to 0^+} \frac{d}{d\beta} f(\rho_G;t=e^{-\beta}) = \frac{\tr_{\textrm{Ad}_G} 1}{2|\Gamma|} \left. \frac{d}{d\beta}\frac{(1-e^{-2\beta})^3}{(1-e^{-3\beta})^2} \right|_{\beta=0}  = \frac{4\dim G}{9|\Gamma|}\ ,
\end{equation}
since $\tr_{\textrm{Ad}_G} 1 =\dim G$. This is \emph{independent} of the gauge holonomy $\rho_G:\Gamma\to G$ and just $\frac{1}{|\Gamma|}$ of the supersymmetric Casimir energy on $S^3$ \cite{Bobev:2015kza} as expected. 
Refining chemical potentials for $SO(6)_R$ R-charges is straightforward and does not affect the independence of the supersymmetric Casimir energy from the gauge holonomy. Therefore, we have found a vacuum degeneracy of four-dimensional $\mathcal{N}{=}4$  super Yang-Mills theory with gauge group $G$ on $S^3/\Gamma$ parametrized by the gauge holonomy $\rho_G:\Gamma \to G$.
Also note that we did not assume anything about $G$ and $\Gamma$, 
so this is  an effect \emph{purely} due to the $\mathcal{N}{=}4$ supersymmetry. 
For general $\mathcal{N}{=}1$ supersymmetric theories, the supersymmetric Casimir energy does depend on the choice of gauge holonomies and turning on a nontrivial holonomy costs energy \cite{Benini:2011nc}.

\section{Duality}
\label{sec:duality}

\subsection{Holographic Chern-Simons dual}
In this subsection, we investigate the vacuum degeneracy of four-dimensional $\mathcal{N}{=}4$ $U(N)$ super Yang-Mills theory on $S^3/\Gamma$, parameterized by $\rho:\Gamma\to U(N)$, from a holographic perspective. (From this point on, we fix the gauge group of 4d $\mathcal{N}{=}4$ super Yang-Mills theory to be $U(N)$.)

Through the AdS/CFT correspondence, the theory is dual to Type IIB string theory on AdS$_5/\Gamma \times S^5$ \cite{Maldacena:1997re}. We shall focus on the low-lying spectrum of this duality when topology of the AdS geometry is given by AdS$_5/\Gamma \cong \bR_t \times B^4/\Gamma$, where $\bR_t$ is the time direction and $B^4$ is a four-dimensional ball \cite{Hawking:1982dh}. Then, the BPS degrees of freedom on this background could be decomposed into propagating degrees in the full ten-dimensional geometry and localized degrees at the fixed point of $\Gamma$. The former will be given by the Kaluza-Klein spectrum of type IIB supergravity on AdS$_5 \times S^5$ \cite{Kinney:2005ej} that are invariant under the action of $\Gamma$. Since this plays no role realizing the vacuum degeneracy, we will focus on the latter. 

Near the fixed point, the geometry behaves like $\bR_t \times \mathbb{R}^4/\Gamma \times S^5$. Type IIB string theory near such a singularity is described by the six-dimensional $\mathcal{N}{=}(2,0)$ superconformal field theory of type $\Gamma$, whose elementary degrees of freedom are given by tensionless self-dual strings originating from D3-branes wrapping the singularity \cite{Witten:1995zh}. In our case, $N$ units of RR 5-form flux $F_5$ are turned on $S^5$. Thus, we have a 6d (2,0) theory coupled to the background 5-form flux on $S^5$.

Although we believe that the above system contains topological degrees of freedom responsible for the huge vacuum degeneracy, such a 6d (2,0) theory on the RR 5-form flux background has not been studied so far, to the knowledge of the authors. Instead, we consider a class of closely-related backgrounds, for which our understanding is more robust, to distill a general lesson. Recall that 4d $\mathcal{N}{=}4$ $U(N)$ super Yang-Mills theory describes the low-energy dynamics of a stack of $N$ M5-branes wrapping a torus $T^2$, whose complex modulus is identified with the complexified gauge coupling $\tau\equiv \frac{\theta}{2\pi} + \frac{4\pi i}{g_\textrm{YM}^2}$ of the $\mathcal{N}{=}4$ super Yang-Mills theory. We shall change the wrapped manifold into a Riemann surface $\Sigma$ of genus $g>1$. Then, the low-energy dynamics of $N$ coincident M5-branes wrapping $\Sigma$ is described by the 4d $\mathcal{N}{=}2$ class S theory of type $U(N)$ \cite{Gaiotto:2009we,Gaiotto:2009hg}. Note that we have included modes corresponding to the center of mass of the M5-brane stack.

The four-dimensional $\mathcal{N}{=}2$ class S theory of type $U(N)$ is holographically dual to M-theory on AdS$_5 \times \Sigma \tildetimes S^4$, where $\Sigma$ is the Riemann surface of genus $g>1$ on which M5-branes are wrapped and $\tildetimes S^4$ means that $S^4$ is nontrivially fibered over $\Sigma$ \cite{Maldacena:2000mw}.%
\footnote{This can be intuitively understood as follows. Recall that the near-horizon geometry of black M5-brane is given by AdS$_7 \times S^4$ with $N$ units of M-theory 4-form flux $G_4$ on $S^4$. Compactifying M5-branes on $\Sigma$ reduces AdS$_7$ to AdS$_5 \times \Sigma$. In addition, to preserve supersymmetries, we perform a topological twist over $\Sigma$ by embedding its $SO(2)$ spin connection into the $SO(5)$ isometry of $S^4$. In this way, squashed $S^4$ is fibered over $\Sigma$ whose topology is roughly $\Sigma \tildetimes S^1 \times S^1 \times S^2$. The 
isometry of the internal manifold is reduced to $U(1) \times SU(2) \subset SO(5)$, which corresponds to the R-symmetry of the dual 4d $\mathcal{N}{=}2$ class S theory descending from that of the 6d $\mathcal{N}{=}(2,0)$ theory living on M5-branes compactified on $\Sigma$.}
Now, let us consider this theory 
on $S^3/\Gamma$, which is dual to M-theory on AdS$_5/\Gamma \times \Sigma \tildetimes S^4$ with $N$ units of M-theory 4-form flux $G_4$ on $S^4$. As before, we focus on the low-lying spectrum of this duality when AdS$_5/\Gamma \cong \bR_t \times B^4/\Gamma$. The BPS degrees of freedom on this background could be decomposed into propagating degrees in the full eleven-dimensional geometry and localized degrees at the fixed point of $\Gamma$. Among them, the latter should account for the nontrivial supersymmetric vacua. Near the fixed point, the geometry behaves like $\bR_t \times \mathbb{R}^4/\Gamma \times \Sigma \tildetimes S^4$. Then, M-theory near such a singularity is described by a seven-dimensional supersymmetric gauge theory, whose gauge group is given by a simply connected compact Lie group $\mathsf{G}$ of ADE type in Table~\ref{tab:McKay} by the McKay correspondence \cite{Sen:1997kz}.

In order to analyze the coupling of the 7d gauge theory to the M-theory flux $G_4$ in detail, let us first consider the case when $\Gamma=\bZ_k$. 
In this case, we regard the transverse space $\bC^2/\bZ_k$ as a Taub-NUT space,
and reduce M-theory along the Taub-NUT circle.
Then, we get type IIA string theory with a stack of $k$ D6-branes wrapping $\bR_t \times \Sigma \tildetimes S^4$. 
In addition, $N$ units of RR 4-form flux $F_4$ are turned on $S^4$. The low-energy dynamics of D6-branes is described by the Dirac-Born-Infeld action and the Wess-Zumino action, together with their supersymmetric completion. Among them, the Wess-Zumino action is responsible for the topological coupling, which includes the following term: 
\begin{equation}\label{WZ}
\begin{aligned}
    &\mu_6 \int_{\bR_t \times \Sigma \tildetimes S^4} \tr \left[ e^{2\pi \alpha'F_{U(k)} +B_2} \wedge \sum_p C_p\right] \\
    \supset \; &\mu_6 \int_{\bR_t \times \Sigma \tildetimes S^4} \tr \left[ \frac{1}{2} (2\pi \alpha' F_{U(k)} +B_2) \wedge(2\pi \alpha' F_{U(k)} +B_2) \wedge C_3\right] \\
    \supset \; &\mu_6 \int_{\bR_t \times \Sigma \tildetimes S^4} 2\pi^2\alpha'^2 \; \tr \left(F_{U(k)} \wedge F_{U(k)} \right)    \wedge C_3 \\
    \supset \;&\mu_6 \int_{\bR_t \times \Sigma} 2\pi^2\alpha'^2 \;\textrm{CS}_{U(k)} \times \int_{S^4} F_4\ ,
    \end{aligned}
\end{equation}
where we applied the integration by parts and the fiber integration successively in the last line.
Here, $\mu_6$ is the D6-brane charge, $F_{U(k)}$ is the field strength of the ${U(k)}$ gauge field living on $k$ coincident D6-branes, $B_2$ is the NS-NS 2-form field, and $F_4=dC_3$ is the RR 4-form flux. Note that the trace is in the $k \times k$ fundamental representation and $\textrm{CS}_{U(k)}$ is the Chern-Simons 3-form of the ${U(k)}$ gauge field normalized as 
\begin{equation}
d\CS_{U(k)}=\tr \left(F_{U(k)} \wedge F_{U(k)} \right)\ .
\label{CSnorm}
\end{equation} 
We disregarded terms irrelevant for the structure of topological couplings.

Next, we fix all normalization coefficients following \cite{Polchinski:1998rr}. The D6-brane charge $\mu_6$ is given by
\begin{equation}
    \mu_6 =  \frac{1}{(2\pi)^6\alpha'^{7/2}}\ ,
\end{equation}
and the RR 4-form flux $F_4$ on $S^4$ is given by
\begin{equation}
    \int_{S^4} F_4 =  N \times \mu_4 (2\kappa_{10}^2) = N \times \frac{(2\pi)^7 \alpha'^4}{(2\pi)^4 \alpha'^{5/2}}\ .
\end{equation}
Integrating out $F_4$ on $S^4$ at the last line of \eqref{WZ}, we get the following low-energy effective action:
\begin{equation}
    \frac{N}{4\pi} \int_{\bR_t \times \Sigma} \textrm{CS}_{U(k)} \ .
    \label{CScoupling}
\end{equation}
Recalling our normalization \eqref{CSnorm},
this is the action of three-dimensional pure $U(k)$ Chern-Simons theory at level $N$ on $\bR_t \times \Sigma$. 
When we go back from the Taub-NUT space to the ALE space,
the $U(1)$ part becomes non-normalizable.
This means that  the $\bC^2/\bZ_k$ singularity 
generates 3d pure $SU(k)$ Chern-Simons theory at level $N$ in M-theory.\footnote{%
At this point, a careful reader might worry that the Chern-Simons level might be shifted 
by integrating out massive fermion fields. Let us argue that this does not happen.
Note that the massive fermion fields in three dimensions arise 
from the Kaluza-Klein expansion of the seven-dimensional fermions along the $S^4$ direction,
and their mass terms are eigenvalues of an appropriate Dirac operator $\slashed{D}$ there.
This anticommutes with $\Gamma:=\gamma^{\mu\nu\rho\sigma} \omega^{\text{vol}(S^4)}_{\mu\nu\rho\sigma}$.
Therefore, all nonzero eigenvalues appear in pairs of the form $\pm m$, and do not shift the Chern-Simons level.
The zero modes give rise to three-dimensional fermions, which are massless in the absence of the Chern-Simons interaction.
We expect that they are the superpartners of the Chern-Simons gauge field,
and that they do not shift the Chern-Simons levels, following the analysis of \cite{Kao:1995gf}.
}

For finite subgroups $\Gamma\subset SU(2)$ other than $\bZ_k$,
we can consider partially blowing up $\bC^2/\Gamma$ to some $\bC^2/\bZ_{k'}$, with $k'\ge 2$.
This corresponds to Higgsing from $\sG$ to $SU(k')$.
As we saw above, we have the level-$N$ Chern-Simons coupling for $SU(k')$.
By consistency, we need to have the level-$N$ Chern-Simons coupling for $\sG$ before the Higgsing.

Finally, we may attempt to apply the above analysis to the case when $\Sigma = T^2$, for which the corresponding class S theory becomes $\mathcal{N}{=}4$ super Yang-Mills theory. However, in such a case, the Maldacena-Nu\~nuez background discussed above becomes singular, preventing a straightforward application. Nevertheless, it is natural to extrapolate the final result. Namely, it seems reasonable to expect that the same Chern-Simons theory, now defined on $T^2$, accounts for the vacuum degeneracy of $\mathcal{N}{=}4$ super Yang-Mills theory on $S^3/\Gamma$.
In what follows, we will show that this expectation is indeed true by explicitly matching the vacuum structures across the duality.

In summary, here we provided an argument that the vacuum degeneracy of four-dimensional $\mathcal{N}{=}4$ $U(N)$ super Yang-Mills theory on $S^3/\Gamma$, parameterized by $\rho:\Gamma\to U(N)$
is the same as that of three-dimensional pure $\mathsf{G}$ Chern-Simons theory at level $N$ on $T^2$. 
Here, $\mathsf{G}$ is the simply connected compact Lie group of ADE type associated with $\Gamma$ by the McKay correspondence, as in Table~\ref{tab:McKay}. Moreover, the complex modulus of $T^2$ equals to the complexified gauge coupling $\tau = \frac{\theta}{2\pi} + \frac{4\pi i}{g_\textrm{YM}^2}$ of the $\mathcal{N}{=}4$ super Yang-Mills theory.%
\footnote{%
We note here again that the duality between 
\Nequals4 super Yang-Mills theory and 3d pure $\sG$ Chern-Simons theory on $T^2$,
and its generalization to more general class S theories on $S^3/\Gamma$
were already provided
in \cite{Ju:2023umb,Ju:2023ssy,Albrychiewicz:2024fkr},
by directly analyzing Euclidean D3-branes or M5-branes on the $\bC^2/\Gamma$ singularity
and studying the effect of the 7d gauge field at the singularity, without taking the holographic dual.
Here we preferred to first go to the holographic dual, for reasons explained 
in the footnote~\ref{foot:1st}.
}\footnote{%
An attempt to relate Type IIB string theory on $\bR_t\times B^4/\Gamma \times S^5$ 
to 3d Chern-Simons theory, which eventually fails, is the following.
Consider $S^5$ as an $S^1$ fibration over $\CP^2$.
If its T-dual along the $S^1$ fiber \emph{were} Type IIA string theory on $\bR_t\times B^4/\Gamma\times \CP^2\times S^1$, 
we \emph{would have} been able to further lift it to M-theory on $\bR_t\times B^4/\Gamma\times \CP^2\times T^2$.
This \emph{would have} a 7d $\sG$ gauge field at the singularity, 
with $N$ units of the $F_4$ flux on $\CP^2$.
Then the same argument as in the general class S case \emph{would have} given us 
3d pure Chern-Simons theory with gauge group $\sG$ and level $N$ on $T^2$.

This cannot be correct, since $\CP^2$ only admits a spin$^\textrm{c}$ structure and not a spin structure.
As M-theory does not have a $U(1)$ symmetry acting on the fermions, 
M-theory cannot be put on $\CP^2$, on which no spin structure exists.
The issue is that the $S^1$ fiber within $S^5$ has a NS spin structure,
whose T-dual yields Type 0A rather than Type IIA \cite{Atick:1988si,Dine:1988nrl},
so that all the fermions in the T-dual comes from the winding modes \cite{Duff:1998us}.
The strongly-coupled limit of Type 0A was studied e.g.~in \cite{Bergman:1999km}
and was claimed to be a certain twisted $S^1$ compactification of M-theory.
However, this cannot be applied in this context, since such a compactification should still
have local fermion fields not coming from wrapped membranes,
meaning that it is incompatible with $\CP^2$.
These issues seem worth investigating further.
} 
We postpone a detailed study of how the vacuum states match across this duality to Sec.~\ref{sec:states},
and move on to the analysis of how the 1-form symmetries are mapped across the duality.

\subsection{On 1-form symmetries}

Let us now discuss how the 1-form symmetries of 4d \Nequals 4 super Yang-Mills theory and the class S theory are implemented
on the side of 3d Chern-Simons theory.
We will follow the same set of steps as we have used before.

\paragraph{1-form symmetries of \Nequals4 super Yang-Mills theory and more general class S theories.}
Four-dimensional \Nequals 4 $U(N)$ super Yang-Mills theory has electric and magnetic $U(1)$ 1-form symmetries.
Let us use $\sB_2$ and $\sC_2$ to denote their background fields.
They are known to have a mixed anomaly given by \begin{equation}
N\int_{5d} \sB_2 \wedge d\sC_2\ . \label{mixed}
\end{equation}

On the holographic side, we can identify them with the NSNS and RR 2-forms, $B_2^\text{IIB}$ and $C_2^\text{IIB}$,
of Type IIB string theory, with a constant wavefunction along $S^5$.
So we can write $B_2^\text{IIB}=\sB_2$ and $C_2^\text{IIB}=\sC_2$.
Then the coupling \eqref{mixed} descends from the Type IIB coupling 
$\propto\int_{10d} F_5 \wedge B_2 \wedge dC_2$ in ten dimensions, with $\int_{S^5} F_5=N$.

Let us next consider the 1-form symmetries of class S theories of type $U(N)$, or equivalently
$N$ M5-branes with the center-of-mass freedom kept inside,
on a Riemann surface $\Sigma$ of genus $g$.
Holographically, we have the Maldacena-Nu\~nuez background as we recalled above,
of the form $M_5\times \Sigma \tildetimes S^4$,
with $\int_{S^4}G=N$.
Therefore, the M-theory coupling \begin{equation}
\propto \int_\text{11d} C\wedge G\wedge G
\end{equation} 
reduces to \begin{equation}
\propto N\int_{M_5\times \Sigma} C\wedge G\ .
\end{equation}
We now take a standard basis $\omega_i$, $\tilde\omega_i$ ($i=1,\ldots,g$) of 1-forms on $\Sigma$ satisfying \begin{equation}
\int_\Sigma \omega_i \wedge \omega_j =0\ , \quad
\int_\Sigma \tilde \omega_i \wedge \tilde\omega_j =0\ , \quad
\int_\Sigma \omega_i \wedge \tilde\omega_j = \delta_{ij}\ ,
\end{equation} and expand the 7-dimensional $C$ field as \begin{equation}
C_3^\text{M}\Bigm|_\text{with legs along $\Sigma$} = 
\sC_2^i \wedge \omega_i + \tilde\sC_2^i \wedge \tilde\omega_i\ .
\label{SigmaC3}
\end{equation}
Then we see that the M-theory coupling further reduces to \begin{equation}
\propto N \int_{5d} \sum_i \sC_2^i \wedge d \tilde \sC_2^i\ , \label{1-form-M}
\end{equation} showing the existence of $g$ pairs of `electric' and `magnetic' $U(1)$ 1-form symmetries.

When $\Sigma=T^2$, the Maldacena-Nu\~nez background is ill-defined.
However, the structure of the 1-form symmetries can also be studied directly
in the system of M5-branes in an M-theory background without taking the holographic limit,
in the more modern SymTFT perspective.
The analysis is essentially unchanged, and now applies also to the case of $T^2$.
Let $x_{1,2}$ are appropriately normalized coordinates of $T^2$
so that we can take $\omega_1=dx_1$ and $\tilde \omega_1= dx_2$.
Further rewriting $(\sB_2,\sC_2):=(\sC_2^1,\tilde \sC_2^1)$, 
we have 
\begin{equation}
C_3^\text{M}\Bigm|_\text{with legs along $T^2$} = \sB_2 \wedge dx_1 +\sC_2 \wedge dx_2\ ,
\label{T2C3}
\end{equation} 
and we see that the anomaly theory \eqref{mixed} of \Nequals4 super Yang-Mills theory
is a special case of \eqref{1-form-M} when $g=1$.

\paragraph{1-form symmetry of the 7d gauge theory.}
We now want 
to know how $\sB_2$ and $\sC_2$ in the case of the torus \eqref{T2C3}, 
or more generally $\sC_2^i$ and $\tilde\sC_2^i$ for a surface $\Sigma$ of genus $g$ \eqref{SigmaC3},  couple to the $\sG$ gauge fields supported on the singularity.
For this, 
we need to consider M-theory on $M_7\times \bR^4/\Gamma$
and study its 1-form symmetry. This was first studied in \cite{Albertini:2020mdx}.
Here we make a brief review of this analysis.
In the following we need several standard facts about the geometry of $S^3/\Gamma$
and the associated ALE space.
For details, see Appendix~\ref{app:geom}.

As it is hard to directly study the singular limit, let us slightly deform the $\bR^4/\Gamma$ singularity
to an ALE space $X_\Gamma$, still with the asymptotic boundary $S^3/\Gamma$.
This corresponds to the situation where the $\sG$ gauge group is broken to its Cartan torus $U(1)^{r_\sG}$
via nonzero vacuum expectation values of the adjoint scalar fields, where $r_\sG=\rank \sG$.

As is well-known, $X_\Gamma$ contains $r_\sG$ two-spheres, $C_1,\ldots,C_{r_\sG}$, corresponding 
to the nodes of the Dynkin diagram of $\sG$. 
They form the basis of a lattice $Q_\sG$ 
and intersect according to the Cartan matrix, meaning that $Q_\sG$ can be identified with the root lattice of $\sG$.

Let $\{D_i\}$ be the 2-cycles  dual to $\{C_i\}$, satisfying  \begin{equation}
\#(C_i \cup D_j) =\delta_{ij}\ .
\end{equation} 
$\{D_j\}$ form a lattice dual to $Q_\sG$. 
Therefore it can be identified with the weight lattice $P_\sG$ of $\sG$,
which contains $Q_\sG$. 
The elements of $P_\sG$ not within $Q_\sG$ are non-compact.
They determine one-cycles on the asymptotic infinity, given by elements of $H_1(S^3/\Gamma;\bZ)$.
In particular, this means that $P_\sG/Q_\sG=H_1(S^3/\Gamma;\bZ)$.

M2-branes wrapped on these non-compact cycles give rise to line operators of the 
seven-dimensional gauge theory. As they are labeled by elements of $P_\sG$,
they are naturally identified as Wilson lines, whose charges are specified by $P_\sG$.
Now, W-bosons of the 7d gauge fields are M2-branes wrapped around compact cycles $C_i$.
So, dynamical particles have charges labeled by $Q_\sG$.
This means that there is a 1-form symmetry associated to the quotient group $P_\sG/Q_\sG$.

For simplicity, assume that $M_7$ is compact, and the only non-compact 
direction of the M-theory spacetime is the radial direction of $X_\Gamma$.
Then, the background field for the above 1-form symmetry is the boundary value of the M-theory 4-form 
on the asymptotic boundary $M_7\times S^3/\Gamma$.
To see this, introduce 2-forms $\alpha_i$ normalized as \begin{equation}
\int_{C_j} \alpha_i=\delta_{ij}\ ,
\end{equation} 
so that $\{\alpha_i\}$ forms a basis of $H^2(X_\Gamma;\bZ)$.
The two-form $\alpha_i$ is the first Chern class of a $U(1)$ bundle $L_i$ on $X_\Gamma$.
The curvature $\alpha_i$ vanishes on its boundary $S^3/\Gamma$,
but  $L_i$ can still be a nontrivial flat $U(1)$ bundle there,
given by a representation $\Gamma\to U(1)$.
Note that the set of such representations $\rho: \Gamma\to U(1)$
is naturally a group, where the group operation is a pointwise product: \begin{equation}
(\rho\rho')(g) :=\rho(g)\rho'(g)\ , \qquad g\in \Gamma\ .
\end{equation}
Let us denote this group by $I_\Gamma$.

Generally, the topological class of a $U(1)$ bundle on a manifold $M$
is classified by its first Chern class, taking values in $H^2(M;\bZ)$.
In the case of $S^3/\Gamma$, the group 
$H^2(S^3/\Gamma;\bZ) $ is known to be equal to $I_\Gamma$.
As is recalled in Appendix~\ref{app:geom},
 it is also nontrivially equal to $P_\sG/Q_\sG$.

Now, recall that the seven-dimensional $U(1)$ fields at the singularity arise as 
components of the M-theory 3-form field of the form \begin{equation}
C_3 = C_3^{(0)}+ \sum_i A_i^\text{7d} \wedge \alpha_i\ ,
\end{equation}
where $A_i^\text{7d}$ are $U(1)$ gauge fields on $M_7$,
and $C_3^{(0)}$ is the background part of the 3-form.\footnote{%
This expression can also be used to derive the Chern-Simons coupling \eqref{CScoupling}
for general $\Gamma$. 
Indeed, plugging this into the M-theory term $\propto\int_{11d} C_3\wedge G_4 \wedge G_4$, we have 
the contribution 
$\sim \sum_{i,j}\int_{M_3} (A_i \wedge dA_j) \int_{X_\Gamma} \alpha_i \wedge \alpha_j \int_{S^4} G_4^{(0)}$, where $M_3 = \bR_t \times \Sigma$.
As the part $\int_{X_\Gamma} \alpha_i \wedge \alpha_j$ provides the Cartan matrix,
and the part $\int_{S^4} G_4^{(0)}$ gives $N$,
we have $\sim N \int_{M_3} C^{ij} A_i \wedge dA_j $, the form of the Chern-Simons term
after restricted to the Cartan subalgebra.
}
The curvature $G_4$ of $C_3$ vanishes 
on the asymptotic boundary of the spacetime $M_7\times S^3/\Gamma$,
but its topological class in $H^4(M_7\times S^3/\Gamma;\bZ)$ does not.

Assuming that the topology of $M_7$ is simple enough, 
we have $H^4(M_7\times S^3/\Gamma;\bZ) \simeq H^2(M_7; P_\sG/Q_\sG)$.
In the low-energy path integral of the M-theory effective action on this spacetime,
we are supposed to \emph{fix} this boundary value, 
and the bulk fields compatible with this boundary value are path-integrated over. 
This is exactly the behavior we expect for a 7d theory, which has a 1-form symmetry with group $P_\sG/Q_\sG$.

So far we performed our analysis in the smooth situation where the 7d gauge group $\sG$ 
is broken to its Cartan torus. 
But the non-Abelian theory also has the 1-form symmetry $Z_\sG$, the center of $\sG$,
which is known to be naturally isomorphic to $P_\sG/Q_\sG$,
the quotient of the weight lattice by the root lattice.
This leads us to conclude that the background field for the $Z_\sG\simeq I_ \Gamma$ 1-form symmetry
of the 7d $\sG$ gauge theory on the singularity 
is also given by the boundary value of the M-theory four-form curvature $G_4$,
taking values in $H^2(M_7;Z_\sG)$.

\paragraph{1-form symmetry of the 3d Chern-Simons theory.}
We can now combine the discussions above and relate the 1-form symmetries
of the 4d theory to the 1-form symmetry of the 3d Chern-Simons theory.
The background fields of two $U(1)$ 1-form symmetries of 4d \Nequals4 $U(N)$ super Yang-Mills theory
are specified by  the boundary values at $S^3/\Gamma$ of two two-form fields $\sB_2$ and $\sC_2$.
In the more general case of class S theory for a genus-$g$ surface $\Sigma$,
what matters is the boundary values at $S^3/\Gamma$ of two-form fields $\sC_2^i$ and $\tilde \sC_2^i$.

In the M-theory dual on $\bR_t\times B^4/\Gamma \times \Sigma \tildetimes S^4$, 
they are encoded in the M-theory three-form \eqref{T2C3} or \eqref{SigmaC3}
specified as the boundary values at $S^3/\Gamma$.
This in turn specifies the background field of the  $Z_\sG$ 1-form symmetry
of the 7d theory on $\bR_t \times \Sigma \tildetimes S^4$.
As $\sC_2^i$ and $\tilde\sC^i_2$ are assumed to be constant along $S^4$,
this can be considered as the background field of the $Z_\sG$ 1-form symmetry 
of the 3d  pure Chern-Simons theory on $\bR_t \times \Sigma$.
This is what we wanted to see.

\paragraph{Mapping of the charged line operators.}
It is also instructive to see how the line operators charged under the 1-form symmetries are mapped across the duality.
The starting point is the 4d \Nequals4 $U(N)$ super Yang-Mills theory.
There, the electric 1-form symmetry couples to a  Wilson loop in the fundamental representation, 
and the magnetic 1-form symmetry couples to an 't Hooft loop of the minimal allowed charge.
As we are interested in the theory on $S^3/\Gamma$,
we consider these line operators wrapped around the 1-cycles of $S^3/\Gamma$,
specified by elements of $H_1(S^3/\Gamma;\bZ)$.
This group is dual to $H^2(S^3/\Gamma;\bZ)=I_\Gamma$ via Poincar\'e duality.
In the holographic dual, they correspond to the fundamental string
and the D-string, wrapped around the 1-cycles of $S^3/\Gamma$
and extending down to the singularity of $B^4/\Gamma$.

In the case of more general class S theory, the 1-forms $\omega_i$ and $\tilde\omega_i$ 
in \eqref{SigmaC3} 
are dual to the 1-cycles $A^i$ and $B^i$ on $\Sigma$, intersecting in the standard manner.
Then the charged line operators in the holographic dual
are M2-branes wrapped on the 1-cycles of $S^3/\Gamma$
extending down to the singularity of $B^4/\Gamma$,
times one of the 1-cycles $A^i$ or $B^i$ of $\Sigma$.
When $\Sigma=T^2$, they reduce to the case of \Nequals4 super Yang-Mills theory,
and fundamental strings and  D-strings  correspond to
M2-branes wrapped around the $x_1$-direction and the $x_2$-direction of $T^2$, respectively.

Finally, from the viewpoint of the 3d Chern-Simons theory with gauge group $\sG$,
M2-branes extending along two directions in $B^4/\Gamma$ to infinity
give rise to very heavy line operators of the theory.
This can be naturally identified with Wilson loops of the $\sG$ Chern-Simons theory,
charged under its $Z_\sG$ 1-form symmetry.
Such line operators can wrap around various directions of $\Sigma$.
In particular, when $\Sigma=T^2$,
the 1-cycle on $T^2$ specifies whether they correspond to 
the electric, magnetic or dyonic $U(1)$ 
1-form symmetry of 4d \Nequals4 $U(N)$ super Yang-Mills theory.

\section{Matching of vacuum structure across duality}
\label{sec:matching}

\subsection{States}
\label{sec:states}
Let us start by matching the structure of the states.
On the side of 4d \Nequals4 $U(N)$ super Yang-Mills theory, the states are labeled by
representations \begin{equation}
\rho: \Gamma\to U(N)\ .
\end{equation}
Each representation can be specified by its irreducible decomposition.
Denoting the irreducible representations of $\Gamma$ by
$\rho_i$ $(i=1,\ldots,\ell_\Gamma)$, where $\ell_\Gamma$ is the number of irreducible representations of $\Gamma$, a general representation $\rho$ is specified by a sequence of non-negative integers $n_i$ 
such that \begin{equation}
\sum_{i=1}^{\ell_\Gamma} n_i \dim\rho_i = N\ .
\label{rhos}
\end{equation}  
Since the trivial one-dimensional representation plays a special role below,
we fix the numbering of $\rho_i$ so that $\rho_{i=1}$ is the trivial representation.

In Table~\ref{tab:diagram}, we list the irreducible representations of $\Gamma\subset SU(2)$
in a graphical manner, following McKay \cite{McKay}.
This was done as follows: we introduce a node for each $\rho_i$, and write $\dim \rho_i$ on its side.
We then draw a line between the nodes for $\rho_i$ and $\rho_j$,
if and only if $\rho_i\otimes V$ contains $\rho_j$,
where $V$ is the standard two-dimensional representation coming from $\Gamma\subset SU(2)$.

Famously as pointed out in \cite{McKay}\footnote{%
According to an article in a Japanese mathematics journal \cite{Matsuzawa},
McKay described the way he found the correspondence as follows: 
\begin{quotation}
It was a stormy night. I wanted to try out the new computer-algebra system I just obtained,
and decided to obtain the eigenvectors of the affine Cartan matrix of type $E_8$,
and lo, the character table of the binary icosahedral group appeared. 
\end{quotation}
Unfortunately this is an English translation of a Japanese translation of McKay's original words,
but it still shows his unique quality: who else would think of trying out the capability of
a new computer-algebra system in this manner,
and then being able to recognize the result as the character table of a particular finite group?
}, 
this procedure gives an extended Dynkin diagram of type $A_\ell$, $D_\ell$ or $E_\ell$,
such that the trivial representation $\rho_{i=1}$ corresponds to the extending node.
Furthermore, this diagram distinguishes $\Gamma\subset SU(2)$,
and  positive integers $\dim\rho_i$ equal the comarks $m_i$, also important in Lie theory.
They are defined as the components of the unique zero-eigenvalue eigenvector of the Cartan matrix,
normalized so that the component $m_{i=1}$ corresponding to the extending node $\rho_{i=1}$ is one.

\def\node#1#2{\overset{#1}{\underset{#2}{\circ}}}
\def\bnode#1#2{\overset{#1}{\underset{#2}{\bullet}}}
\def\ver#1#2{\overset{{\llap{$\scriptstyle#1$}\displaystyle\circ{\rlap{$\scriptstyle#2$}}}}{\scriptstyle\vert}}
\begin{table}
\[
\begin{array}{c@{\quad}c@{\quad}c}
\text{Type} & \text{Diagram} & \text{Outer automorphism action} \\
\hline
A_{k-1}  & 
\begin{array}{c}
\raisebox{-12pt}{\rotatebox{30}{$-\!\!-\!\!-$}}\bnode{}{1}\raisebox{0pt}{\rotatebox{-30}{$-\!\!-\!\!-$}} \\[-7pt]
\node{}{1}-\node{}{1}-\cdots-\node{}{1} \\
\end{array}
&\text{$\bZ_k$ rotation}\\[1.5em]
D_{k+2}  & \bnode{}{1}-\node{\ver{}{1}}{2}-\node{}{2}-\cdots-\node{\ver{}{1}}{2}-\node{}{1} &
\begin{array}{cl}
\bZ_4& \text{($k$: odd)}  \\
\bZ_2\times \bZ_2& \text{($k$: even)}
\end{array}
 \\[.8em]
E_6  & \bnode{}{1}-\node{}{2}-\node{\overset{\ver{}{1}}{\ver{}{2}}}{3}-\node{}{2}-\node{}{1} 
&\text{$\bZ_3$ rotation}\\[1em]
E_7 & \bnode{}{1}-\node{}{2}-\node{}{3}-\node{\ver{}{2}}{4}-\node{}{3}-\node{}{2}-\node{}{1}
&\text{$\bZ_2$ flip}\\[1em]
E_8 & \bnode{}{1}-\node{}{2}-\node{}{3}-\node{}{4}-\node{}{5}-\node{\ver{}{3}}{6}-\node{}{4}-\node{}{2} 
&\text{no nontrivial action}
\end{array}
\]
\caption{Extended Dynkin diagrams of type ADE, and the corresponding outer automorphism actions.
The integers at the nodes are comarks $m_i$ in the Lie theory,
while they are the dimensions $\dim\rho_i$ of irreducible representations of $\Gamma$.
The extending node is filled in black.
\label{tab:diagram}}
\end{table}

In contrast, the states of pure Chern-Simons theory with gauge group $\sG$ and level $N$ on $T^2$ are
well-known to be labeled by the irreducible representations of affine Lie algebra of type $\sG$ with level $N$.
An irreducible representation  is specified by an affine weight vector \begin{equation}
\lambda = \sum_{i=1}^{r_\sG+1} n_i \pi_i\ ,
\label{affine-weights}
\end{equation} where $\pi_i$'s are the basis elements of the weight lattice
and $n_i$'s are non-negative integers, such that \begin{equation}
\sum_{i=1}^{r_\sG+1} n_i m_i= N\ .
\label{lambdas}
\end{equation}
We take the convention that $\pi_{i=1}$ is the extending node, corresponding to the lowest root of the root lattice.

The basic McKay correspondence is $m_i=\dim \rho_i$ (with $\ell_\Gamma = r_\sG +1$),
and therefore we have a nice one-to-one correspondence between the vacuum states of
four-dimensional \Nequals4 super Yang-Mills theory with gauge group $U(N)$ on $S^3/\Gamma$, 
specified by \eqref{rhos},
and the vacuum states of three-dimensional pure Chern-Simons theory with gauge group $\sG$ at level $N$ on $T^2$,
specified by \eqref{lambdas}.

This correspondence had been noticed previously.  
In mathematics, this observation at least goes back to \cite{Nakajima:1994nid},
and the earliest mention on the string theory side is \cite{Vafa:1994tf} citing it.
The string duality explaining this correspondence was first given in \cite{Ju:2023umb},
to the authors' knowledge.

Before moving on, note that the number of solutions to \eqref{rhos}, or equivalently \eqref{lambdas},
scales polynomially at large $N$ as
\begin{equation}
\sim N^{\ell_\Gamma-1}=N^{r_\sG}\ .
\label{poly-deg}
\end{equation}
In the large $N$ limit, where the AdS/CFT correspondence is usually considered,
this is a huge degeneracy, much larger than any degeneracy controlled by the group theory of  $I_\Gamma=Z_\sG$ which is independent of $N$.
But this polynomial degeneracy \eqref{poly-deg} is still much smaller than the degeneracy of black hole microstates,
which grows exponentially with $N^2$.

\subsection{Action of 1-form symmetries}

\subsubsection{\Nequals4 super Yang-Mills theory}
\label{subsec:4d-1-form}

Let us now discuss the action of 1-form symmetries. 
We start from the side of \Nequals4 $U(N)$ super Yang-Mills theory,
which has electric and magnetic $U(1)$ 1-form symmetries.

\paragraph{The electric one-form symmetry.}
The electric $U(1)$ one-form symmetry of a $U(N)$ gauge theory
acts by shifting a given $U(N)$ gauge field $A_\mu^{U(N)}$
by adding a $U(1)$ gauge field $A_\mu^{U(1)}$ via \begin{equation}
A_\mu^{U(N)} \mapsto (A_\mu^{U(N)})' := A_\mu^{U(N)} + A_\mu^{U(1)} \mathbf{1}_{N\times N}\ .
\end{equation}
In our case, a $U(1)$ gauge field on $S^3/\Gamma$ 
is specified by a 1-dimensional representation \begin{equation}
\sigma: \Gamma\to U(1)\ .
\end{equation}
They correspond to nodes of the extended Dynkin diagram filled in black and labeled by $1$.
Let us then introduce a notation \begin{equation}
I_\Gamma:=\{\sigma:\Gamma\to U(1)\}\ ,
\end{equation}
which form a group under the pointwise multiplication, i.e.~\begin{equation}
(\sigma\sigma')(g) := \sigma(g)\sigma'(g)\ ,\qquad g\in \Gamma\ .
\end{equation}
We can equally say that we take the tensor product.
This is known to agree with $H^2(S^3/\Gamma;\bZ)$.

The group $I_\Gamma$ acts on the vacua labeled by $\rho:\Gamma\to U(N)$ 
again by a pointwise product, \begin{equation}
(\sigma\rho)(g) := \sigma(g)\rho(g)\ ,
\end{equation} or equivalently via the tensor product.
On the ket vector, we can describe its action by the formula  \begin{equation}
\u\sigma \ket{\rho} := \ket{\sigma\rho}\ .
\end{equation}
Here we used $\u\sigma$ to denote the action of the symmetry group
on the vacuum states.

In terms of the labels $n_i$ satisfying $\sum_{i=1}^{\ell_\Gamma} n_i\dim\rho_i=N$,
its action can be expressed as follows. 
As $\sigma\rho_i$ is also an irreducible representation,
we can define $\sigma(i)$ by the equation \begin{equation}
\rho_{\sigma(i)} = \sigma\rho_i\ . 
\end{equation}
Let us write a general representation as $\rho=\sum_i n_i\rho_i$.
Then the action of $\sigma$  is given by \begin{equation}
\u\sigma\ket{\bigoplus_i n_i\rho_i} = \ket{\bigoplus_i n_i\rho_{\sigma(i)}}\ .
\label{electric-action}
\end{equation}

\paragraph{The magnetic one-form symmetry.}
Let us move on to the magnetic one-form symmetry.
The magnetic $U(1)$ one-form symmetry of the $U(N)$ gauge theory
on a spatial manifold $M_3$
is  determined by the topological information
contained in its first Chern class, $c_1(A)\in H^2(M_3;\bZ)$,
of a classical $U(N)$ gauge configuration $A$.
The group $H^2(M_3;\bZ)$ is the Abelian group of the \emph{charges} under the magnetic $U(1)$ symmetry.
Therefore, the symmetry \emph{group} is its Pontryagin dual,
and an element in it is a homomorphism
\begin{equation}
\omega: H^2(M_3;\bZ)\to U(1)\ ,
\end{equation} and it acts on the ket $\ket{A}$ via \begin{equation}
\u\omega\ket{A} := \omega(c_1(A)) \ket{A}\ .
\end{equation}
Here we used $\u\omega$ to denote the action of the symmetry group
on the vacuum states, as before.

In our case, $H^2(M_3;\bZ)=H^2(S^3/\Gamma;\bZ)=I_\Gamma$,
so a $U(N)$ bundle $A$ should define an element $c_1(A)\in I_\Gamma$,
and the magnetic $U(1)$ 1-form symmetry acts via an action of $\hat I_\Gamma$,
the Pontryagin dual of $I_\Gamma$.
Now, a flat $U(N)$ bundle specified by $\rho: \Gamma\to U(N)$
defines a flat $U(1)$ bundle specified by $\det\rho:\Gamma\to U(1)$,
which can naturally be considered as an element  of
$I_\Gamma$.
This is known to be the first Chern class $c_1$ of this flat $U(N)$ bundle.
Therefore, an element $\omega: I_\Gamma \to U(1)$ acts on the vacuum state $\ket{\rho}$
via \begin{equation}
\u\omega \ket{\rho} = \omega(\det\rho) \ket{\rho}\ .
\end{equation}

More explicitly, note $\det\rho_i\in I_\Gamma$ for each $\rho_i$.
Then \begin{equation}
\det\rho =\prod_{i=1}^{\ell_\Gamma}(\det\rho_i)^{n_i} \in I_\Gamma\ ,
\end{equation} for $\rho=\bigoplus_i n_i\rho_i$. 
We therefore have the action \begin{equation}
\u\omega\ket{\bigoplus_i n_i\rho_i} = \prod_i \omega(\det\rho_i)^{n_i} \ket{\bigoplus n_i\rho_i}\ .
\label{magnetic-action}
\end{equation}

\paragraph{The mixed anomaly.}
We have seen that the electric $U(1)$ 1-form symmetry acts via $I_\Gamma$
and the magnetic $U(1)$ 1-form symmetry acts via $\hat I_\Gamma$.
These actions do not necessarily commute. Indeed, it is easy to see that \begin{equation}
\u\omega \u\sigma \ket{\rho} = \omega(\det(\sigma\rho))\ket{\sigma\rho}
=\omega(\sigma)^N \omega(\det\rho) \ket{\sigma\rho}
=\omega(\sigma)^N \u\sigma\u\omega\ket{\rho}\ ,
\end{equation}that is, \begin{equation}
\u\omega\u\sigma =\omega(\sigma)^N \u\sigma\u\omega\ .
\label{omega-sigma-anomaly}
\end{equation}
This is in accord with the mixed anomaly between electric and magnetic $U(1)$ 1-form symmetries,
which is proportional to $N$.

As a concrete example, consider the $U(N)$ theory on $S^3/\bZ_k$. 
Then $I_\Gamma$ and $\hat I_\Gamma$ are both $\bZ_k$.
Denoting their generators by $\sigma$ and $\omega$, 
we have $\omega(\sigma)=e^{2\pi i/k}$, so we find \begin{equation}
\u{\omega}\u{\sigma}= e^{2\pi i N/k} \u{\sigma}\u{\omega}\ .
\end{equation}
This means that the actions of $I_\Gamma$ and $\hat I_\Gamma$
do commute when $N$ is a multiple of $k$.
The mixed anomaly between electric and magnetic $U(1)$ 1-form symmetries is still there;
it is just that it cannot be seen in the action on the vacuum states on $S^3/\Gamma$.

\subsubsection{Pure Chern-Simons theory}

\paragraph{Generalities.}
Let us now discuss the 1-form symmetry of 3d pure Chern-Simons theory.
For $\mathfrak{su}(N)_k$, it was already taken up as one of the important examples 
in the original paper of higher-form symmetries, see \cite[Sec.~4.4]{Gaiotto:2014kfa},
and a more detailed study was given in \cite{Hsin:2018vcg}.
For general $\mathfrak{g}_k$, it was summarized in  \cite[Sec.~3.1 and Appendix~B]{Closset:2025lqt}.
This is also very much related and almost equivalent to what had been known for decades
under the name of the outer automorphism action of affine Lie algebras, 
as can be found in \cite[Sec.~14.2, 14.6.4]{DiFrancesco:1997nk} and will be recalled below.

We take its gauge group $\sG$ to be compact, simply connected, and non-Abelian.
It has an electric 1-form symmetry group $Z_\sG$ given by the (necessarily finite) center of $\sG$.
Its background field on a 3-manifold $M_3$ is then a cohomology class $w_2\in H^2(M_3;Z_\sG)$,
measuring the obstruction of a $\sG/Z_\sG$ gauge field to be lifted to a genuine $\sG$ gauge field.

The Chern-Simons level $N$ is normalized 
so that $Nd\CS_\sG=N\tr (F_\sG \wedge F_\sG)$ integrates to an integer on closed 4-manifolds
for genuine $\sG$ gauge fields. But this is not guaranteed when the background field $w_2$ 
for the electric 1-form symmetry is nonzero, so that the gauge field to be well-defined only 
as $\sG/Z_\sG$ gauge field. 
This means that the 1-form symmetry has an anomaly proportional to $N$.

\paragraph{Using the technique of topological field theories.}
The discussion up to this point applies to any gauge theory with only fields in the adjoint representation.
But for pure Chern-Simons theory, it is useful to use the technique of topological field theories
to analyze the properties of the 1-form symmetry.
What we do is to shrink as much as possible the region of the 3-manifold 
in which the background field of the 1-form symmetry is supported. 
Such a line operator, labeled by an element $z\in Z_\sG$,
is characterized by the fact that there is a holonomy $z\in Z_\sG\subset \sG$
around a small circle surrounding it.

\begin{figure}[h]
\centering
\begin{tikzpicture}[scale=1.7]
\draw (0,0) ellipse (1.6 and .9);
\draw[red] (0,0) ellipse (1.2 and .6);
\node[red] at (-1.3,0){$\lambda$};
\begin{scope}[scale=.8]
\draw[rounded corners=48pt] (-1.1,.1)--(0,-.6)--(1.1,.1);
\draw[rounded corners=36pt] (-.9,0)--(0,.6)--(.9,0);
\end{scope}
\draw[fill,white] (.2,-.6) circle (.1);
\draw[densely dashed] (0,-.9) arc (270:90:.2 and .365) ;
\draw (0,-.9) arc (-90:90:.2 and .365);
\node at (0.45,-.35) {$\mathcal{B}(z)$}  ;
\end{tikzpicture}
\caption{The line operator $\cB(z)$ acting on the state $\ket{\lambda}$.\label{fig:tikz}}
\end{figure}

Let us now consider a torus $T^2$ with the A-cycle and the B-cycle specified,
and denote the action of this line operator wrapped around the B-cycle
by $\cB(z)$.
As the basis states of the system, we consider a solid torus where the B-cycle is shrunk,
with a line operator $\lambda$ is inserted in it. 
Let us denote by $\ket{\lambda}$ the state prepared by it.
See Figure~\ref{fig:tikz}.
We should then have the action \begin{equation}
\cB(z) \ket{\lambda} = z(\lambda) \ket{\lambda}\ ,
\label{how-z-should-act}
\end{equation}
where $z(\lambda)\in U(1)$  is the scalar by which the element $z\in Z_\sG$
acts in the irreducible representation with weight $\lambda$.
Note that $z(\lambda)=\prod_i z(\pi_i)^{n_i}$ when $\lambda=\sum_i n_i \pi_i$.

Any line operator of the Chern-Simons theory is labeled by an affine weight 
satisfying the constraint \eqref{lambdas}.
Therefore $\cB(z)$ should equal to $B(\mu_z)$ for some affine weight $\mu_z$.
To identify such $\mu_z$, we need to recall a few facts about affine Lie algebras.

An affine Lie algebra of type $\sG$ has an outer automorphism group $O_\sG$,
which acts on the extended Dynkin diagram by a symmetry of the diagram.
Let us say that $\pi_i$ is mapped to $\pi_{o(i)}$ by an element $o\in O_\sG$.
Then, the action is such that $o(1)\neq o'(1)$  if $o\neq o'$. 
Furthermore, $O_\sG$ is abstractly isomorphic to $Z_\sG$.
These information makes it fairly easy to identify the action of $O_\sG$ on the extended Dynkin diagram, 
except possibly for type $D$.
Namely, it is a $\bZ_k$ rotation for $A_{k-1}$,
a $\bZ_3$ rotation for $E_6$, a $\bZ_2$ flip for $E_7$, and no nontrivial operation for $E_8$.
Finally for type $D_{k+2}$, it is $\bZ_4$ or $\bZ_2\times \bZ_2$ depending on the parity of $k$, as summarized in Tab. \ref{tab:diagram}.

The isomorphism of $O_\sG$ and $Z_\sG$ can be made more concrete in the following manner. 
An element $o\in O_\sG$ maps the node $i=1$ to some other node $o(1)$,
such that $m_{o(1)}=1$. Recall that the node $i=1$ corresponded to the trivial representation $\rho_{i=1}$.
Then $\pi_{o(1)}$ is an affine weight vector, and defines an element in $P_\sG$,
which can be projected to $P_\sG/Q_\sG\simeq Z_\sG$.
This map \begin{equation}
O_\sG \ni o \mapsto z_o \in Z_\sG
\label{zo}
\end{equation} is the isomorphism we wanted to describe.
The most important equation for us is the relation \begin{equation}
S_{o(\mu)\lambda} = S_{\mu\lambda}  z_o(\lambda)\ ,
\label{crucial}
\end{equation} 
which is Eq.~(14.255) of \cite[Sec.~14.6.4]{DiFrancesco:1997nk} in our notation,
where $S_{\mu\lambda}$ is the modular $S$-matrix
of Chern-Simons theory.
This relation was proved there using an explicit expression of $S_{\mu\lambda}$.

In general in a 3d topological theory, 
a line operator with the label $\mu$ wrapped on the B-cycle
is known to act on the states on $T^2$ as \begin{equation}
B(\mu)\ket\lambda = \frac{S_{\mu \lambda}}{S_{\One \lambda}}\ket{\lambda}\ ,
\label{general-B-action}
\end{equation}
where $\One$ is the identity line operator of the theory.
From \eqref{crucial}, we then have \begin{equation}
B(o(\One))\ket\lambda = z_o(\lambda)\ket\lambda\ .
\label{B-action}
\end{equation}
This is exactly the relation \eqref{how-z-should-act}, therefore we have
$B(o(\One))=\cB(z_o)$.
In other words, the line operator representing the 1-form symmetry action 
specified by $z\in Z_\sG$ is labeled by the affine weight $o_z(\One)$,
where $Z_\sG\ni z \mapsto o_z\in O_\sG$ is the inverse isomorphism to \eqref{zo}.

Let us next determine how $\cA(z_o)=A(o(\One))$, i.e.~the line operator $o(\One)$
wrapped around the A-cycle,  acts on the state.
We have $A(\mu)=S^{-1} B(\mu) S$ where $S$ is the modular $S$-matrix.
We take the convention that \begin{equation}
S\ket\lambda=\sum_\mu \ket\mu S_{\mu\lambda}\ .
\end{equation} Then we have \begin{equation}
\begin{aligned}
\cA(z_o)\ket\lambda= A(o(\One)) \ket{\lambda} 
&= \sum_{\lambda',\mu }\ket{\lambda'} \overline{S_{\lambda' \mu}} z_o(\mu)    S_{\mu \lambda}  \\
&= \sum_{\lambda',\mu }\ket{\lambda'} \overline{S_{o^{-1}(\lambda') \mu}}  S_{\mu \lambda} 
= \sum_{\lambda'} \ket{\lambda'}\delta_{o^{-1}(\lambda') \lambda} 
= \ket{o(\lambda)}\ .
\end{aligned}
\label{A-action}
\end{equation}

\paragraph{Anomaly.}
Note that $\cA(z_o)=A(o(\One))$ and $\cB(z_{o'})=B(o'(\One))$ do not necessarily commute. 
A short computation reveals that \begin{equation}
B(o'(\One))A(o(\One)) = z_{o'}(o(\One)) A(o(\One))B(o'(\One))\ .
\end{equation}
Now, the identity line operator $\One$ corresponds to the affine weight $N\pi_1$
in the Chern-Simons theory with gauge group $\sG$ at level $N$.
Therefore, $o(\One)=N\pi_{o(1)}$, and we have \begin{equation}
B(o'(\One))A(o(\One)) = z_{o'}(\pi_{o(1)})^N  A(o(\One))B(o'(\One))\ ,
\end{equation}
showing the projective phase due to the anomaly of the 1-form symmetry,
which is proportional to $N$.
This should be compared to \eqref{omega-sigma-anomaly}.

As a concrete example, let us consider the case of $\sG=SU(k)$, for which $Z_\sG=\bZ_k$. Let $o$ denote the generator. 
This rotates the extended Dynkin diagram cyclically by one unit.
Then we find \begin{equation}
B(o(\One))A(o(\One))=z_{o}(\pi_{o(1)})^N A(o(\One))B(o(\One))\ .
\end{equation}
Here $z_{o}(\pi_{o(1)})$ is the eigenvalue by which the central element $o\in \bZ_k\subset SU(k)$ 
acts on the representation specified by $\pi_{o(1)}$, i.e.~the fundamental $k$-dimensional representation. This is simply $e^{2\pi i/k}$.
Therefore we have \begin{equation}
B(o(\One))A(o(\One))=e^{2\pi i N/k} A(o(\One))B(o(\One))\ .
\end{equation}
Note that the two operators commute when $N$ is a multiple of $k$.
This is as it should be, since the anomaly of the 1-form symmetry of the $SU(k)$ Chern-Simons theory is $N$ mod $k$.

\subsection{Agreement of 1-form symmetry actions}
\label{sec:1-form-matching}

We have seen that the actions of the electric and magnetic $U(1)$ 1-form symmetries 
on the vacuum states of \Nequals4 $U(N)$ super Yang-Mills theory on $S^3/\Gamma$
are the projective action of $I_\Gamma\times \hat I_\Gamma$ 
given by \eqref{electric-action} and \eqref{magnetic-action}.
In contrast,  the actions of the $Z_\sG$ 1-form symmetry on the vacuum states on $T^2$ of
pure Chern-Simons theory of gauge group $\sG$ and level $N$ 
are the projective action of $Z_\sG\times Z_\sG$  given  by \eqref{A-action} and \eqref{B-action}.
They look very similar, as a quick inspection of the equations shows.
Let us be more precise. 
\begin{itemize}
\item 
On the electric side, we first need a natural identification $I_\Gamma \simeq Z_\sG$.
To get one, recall that $z\in Z_\sG$
can be naturally identified with $o_z\in O_\sG$, 
which is then naturally identified with
the image $o(i)$ of the node $i=1$ of the extended Dynkin diagram.
As $m_{o(i)}=1$, $\dim\rho_{o(i)}=1$, and therefore  $o(i)\in I_\Gamma$.
This gives a natural identification between $I_\Gamma \simeq Z_\sG$. 

For $\sigma_o=o(1)\in I_\Gamma$,
we then need to have  \begin{equation}
\sigma_o \otimes \rho_i = \rho_{o(i)}\ .
\label{E-nontrivial}
\end{equation} 
This is nontrivial, as 
the left hand side is defined via the tensor product of one-dimensional representations of $\Gamma$,
while the right hand side is defined via the outer automorphism action of the affine Lie algebra of type $\sG$.
\item 
On the magnetic side, we first need a natural identification $\hat I_\Gamma \simeq Z_\sG$.
In other words, we need a natural pairing $Z_\sG \times I_\Gamma \to U(1)$.
One such pairing is given by \begin{equation}
Z_\sG\times I_\Gamma \ni (z,o) \mapsto z( o(\pi_1) )\in U(1)\ ,
\end{equation}
which turns out to be perfect.
This gives another natural correspondence $Z_\sG\ni z \mapsto \omega_z\in \hat I_\Gamma$.
Then we need the agreement \begin{equation}
\omega_z(\det\rho_i) = z (\pi_i)\ .
\label{M-nontrivial}
\end{equation}
Again this is nontrivial, since the left hand side is defined in terms of $\det$ of the representations of $\Gamma$,
while the right hand side is defined directly by the action of the center $Z_\sG$ of $\sG$.
\end{itemize}
We can check the two agreements \eqref{E-nontrivial} and \eqref{M-nontrivial}
by straightforward but tedious case-by-case inspections.
The details are provided in Appendix~\ref{sec:checks}.

\subsection{An application}
Let us discuss two minor applications of the analysis we have done so far.
\paragraph{S-duality of two $U(1)$ 1-form symmetries.}
In Sec.~\ref{subsec:4d-1-form}, we described 
the actions of the electric and magnetic $U(1)$ 1-form symmetries of \Nequals4 super Yang-Mills theory
on the Hilbert space of states on $S^3/\Gamma$.
The electric action was given in \eqref{electric-action},
and the magnetic one was given in \eqref{magnetic-action}.
The S-duality of \Nequals4 $U(N)$ super Yang-Mills theory 
should exchange them, 
but this is not at all apparent in the  analysis  in Sec.~\ref{subsec:4d-1-form}
purely on the side of \Nequals4 super Yang-Mills theory.

However, this becomes perfectly clear after passing to the Chern-Simons description,
since the two actions correspond to \eqref{A-action} and \eqref{B-action}.
Namely, they are given by the same set of invertible line operators,
but wrapped around the $A$-cycle and the $B$-cycle of $T^2$.
They are clearly exchanged by the matrix $S$ of the $SL(2,\bZ)$ action 
associated to the Chern-Simons theory.

This gives us an additional information that 
the S-duality of \Nequals4 $U(N)$ super Yang-Mills theory on $S^3/\Gamma$
should act on the vacuum states $\ket{\rho}$  via the matrix $S$ of 
the pure Chern-Simons theory with gauge group $\sG$ and level $N$.
This is a very natural but surprising prediction.
It would be nice if there is a way to check this on the side of \Nequals4 super Yang-Mills theory,
say by supersymmetric localization.

\paragraph{S-duality of $SU(N)$ and $SU(N)/\bZ_N$.}
So far we have been only discussing the self-S-dual case of \Nequals4 $U(N)$ super Yang-Mills theory.
As another application of our discussions so far, let us perform a small check of
S-duality of \Nequals4 super Yang-Mills theory with gauge group $SU(N)$ and $SU(N)/\bZ_N$,
by comparing the vacuum degeneracy on $S^3/\Gamma$.

Let us start with $SU(N)$. Vacuum states are labeled by  homomorphisms \begin{equation}
\rho: \Gamma\to SU(N)\ .
\end{equation}
This is equivalent to say that they are labeled by  homomorphisms \begin{equation}
\rho:\Gamma\to U(N)\ ,
\end{equation} such that \begin{equation}
\det\rho=1\ .
\end{equation}
In view of \eqref{magnetic-action}, this is equivalent to the condition that \begin{equation}
\u{\omega} \ket{\rho}=\ket \rho\ , \quad\text{for all}\quad \omega\in \hat I_\Gamma\ .
\end{equation}
In other words, the Hilbert space $\cH_{SU(N)}$ of the $SU(N)$ theory on $S^3/\Gamma$
is the subspace of the Hilbert space $\cH_{U(N)}$ of the $U(N)$ theory on $S^3/\Gamma$
invariant under the action of the magnetic $U(1)$ symmetry.
That is, we have \begin{equation}
\cH_{SU(N)} = \left(\frac{1}{|\hat I_\Gamma|}\sum_{\omega\in \hat I_\Gamma} \u{\omega} \right)\cH_{U(N)}\ .
\end{equation}

Next, let us consider $SU(N)/\bZ_N$. Vacuum states are labeled by homomorphisms \begin{equation}
\tilde\rho: \Gamma\to SU(N)/\bZ_N\ .
\end{equation} 
As $SU(N)/\bZ_N=U(N)/U(1)$, this is equivalent to saying that they are labeled by 
projective representations of $\Gamma$.
As $H^2(B\Gamma;U(1))=0$ for finite subgroups $\Gamma$ of $SU(2)$, 
there are no genuine projective representations of $\Gamma$.\footnote{%
This vanishing of $H^2(B\Gamma;U(1))$ can be checked by direct computation.
A geometric proof is provided at the end of Appendix~\ref{app:geom1}.
}
This means that there is a genuine representation \begin{equation}
\rho:\Gamma\to U(N)\ ,
\end{equation} whose reduction to $U(N)/U(1)=SU(N)/\bZ_N$ agrees with $\tilde\rho$.
Such $\rho$ are, however, not generically unique.
Suppose  $\rho$ and $\rho'$ both reduce to the same $\tilde\rho$.
This means that there is a map \begin{equation}
\sigma:\Gamma\to U(1)\ ,
\end{equation} such that \begin{equation}
\rho'(g) =\sigma(g)\rho(g)\ ,
\end{equation} for all $g$. 
By inspection we see that $\sigma$ is actually a representation.
Therefore, $\rho$ and $\rho'$ are related 
by the action  \eqref{electric-action} of the electric 1-form symmetry.
We can then represent a state for $\tilde{\rho}$ as the average over all such $\rho$, by defining \begin{equation}
\ket{\tilde\rho} := \sum_{\sigma\in I_\Gamma}\u{\sigma} \ket{\rho}\ .
\end{equation}
Thus, we see that the Hilbert space $\cH_{SU(N)/\bZ_N}$ of the $SU(N)/\bZ_N$ theory
can be written as \begin{equation}
\cH_{SU(N)/\bZ_N} = \left(\frac{1}{|I_\Gamma|}\sum_{\sigma\in I_\Gamma} \u{\sigma} \right)\cH_{U(N)}\ .
\end{equation}

We now invoke our result that the actions of the magnetic and electric $U(1)$ symmetries
on the states $\ket\rho$ are conjugate by the matrix $S$ of the corresponding
pure Chern-Simons theory. We immediately have \begin{equation}
\cH_{SU(N)/\bZ_N}= S\cH_{SU(N)}\ .
\end{equation} In particular, \begin{equation}
\dim \cH_{SU(N)/\bZ_N}= \dim \cH_{SU(N)}\ ,
\end{equation} meaning that the number of holonomies $\Gamma\to SU(N)$ 
and the number of holonomies $\Gamma\to SU(N)/\bZ_N$ are equal to each other.

\section{Conclusions and discussions}
\label{sec:conclusions}

In this paper, we first established the vanishing of the supersymmetric Casimir energy of
\Nequals4 super Yang-Mills theory on $S^3/\Gamma$,
uniformly for all gauge group $G$, the finite subgroup $\Gamma$ of $SU(2)\simeq S^3$,
and the holonomy sector specified by $\rho_G:\Gamma\to G$.
We then focused on the case when $G=U(N)$
and considered its holographic dual, which contains an ADE singularity in Type IIB string theory.
This led to the finding that the 6d (2,0) theory of type $\sG$, when put on $S^5$
with $N$ units of $F_5$ flux, has a large vacuum degeneracy which grows polynomially with $N$.
Here, $\sG$ is determined by $\Gamma$ via the McKay correspondence.

Unfortunately, this system itself is somewhat difficult to analyze, so we turned to
the holographic dual of the class S theories of type $U(N)$, defined on a Riemann surface $\Sigma$ 
of genus $g>1$. 
Its holographic dual contains an ADE singularity in M-theory,
from which we could see that there is a vacuum degeneracy given by 3d
pure Chern-Simons theory with gauge group $\sG$ and level $N$ on $\Sigma$.
We also paid close attention to how the 1-form symmetries of the class S theories
are mapped to the 1-form symmetry of the Chern-Simons theory.

Assuming that this mapping to the Chern-Simons theory works for the case when $\Sigma=T^2$,
for which the class S theory reduces to \Nequals4 super Yang-Mills theory,
we compared the structure of the vacuum degeneracy of the \Nequals4 super Yang-Mills theory on $S^3/\Gamma$
and that of the pure Chern-Simons theory of gauge group $\sG$
and level $N$ on $T^2$.
We found that the actions of the 1-form symmetry were consistent across this duality,
where the S-duality of \Nequals4 super Yang-Mills theory is mapped to the S-transformation of the torus $T^2$.

We note that all that we discussed in this paper are heavily influenced and indebted to the works \cite{Ju:2023umb,Ju:2023ssy,Albrychiewicz:2024fkr}.
The difference compared to these works are that 
(i) we established the vanishing of the \Nequals4 supersymmetric Casimir energy in a uniform fashion for all cases,
(ii) we preferred to use holography rather than directly analyzing D3 or M5-branes hitting $\bC^2/\Gamma$ singularity, and 
(iii) we discussed the 1-form symmetry on both sides of the duality.

There are a few future directions we can envisage.
Purely field theoretically, we can study the vacuum degeneracy on $S^3/\Gamma$ 
of \Nequals4 super Yang-Mills theories with S-dual gauge groups $G$ and $G^\vee$.
We have a basic prediction that the number of homomorphisms $\Gamma\to G$ 
and the number of homomorphisms $\Gamma\to G^\vee$ should agree.
This is a nontrivial mathematical statement,
which has been checked for various choices of $(G,\Gamma)$ in \cite{Kojima:2025hzr},
not just for $G=U(N)$, $SU(N)$ and $SU(N)/\bZ_N$ studied in this paper.
It will be worthwhile to extend this analysis further.
We can also study not just the vacuum degeneracy, but the entirety of the supersymmetric index,
and try to show its equality under S-duality. 
It would also be nice to study the action of S-duality on the vacuum states.
Our analysis strongly suggests that it is given by the matrix $S$ of the corresponding Chern-Simons theory,
but is there a way to confirm this purely on the side of \Nequals4 super Yang-Mills theory?

Holographically, we can try to extend the study to systems with fewer supersymmetries. 
We already heavily borrowed the general ideas of the paper \cite{Albrychiewicz:2024fkr},
where the study of the vacuum degeneracy of the class S theories of type $A$ on $S^3/\Gamma$ for some $\Gamma$ was initiated.
Further-field theoretical and holographic studies on this question would be worthwhile.
We can also try to extend the Type IIB analysis from AdS$_5/\Gamma\times S^5$ to more general AdS$_5/\Gamma \times X_5$, where $X_5$ is one of \Nequals2 or \Nequals1 orbifolds $S^5/\Gamma'$, or the Sasaki-Einstein manifolds $Y^{p,q}$.
It might also be interesting to replace the gauge group of the \Nequals4 super Yang-Mills theory from 
$u(N)$ to $so(2N)$, $so(2N+1)$ and $sp(N)$. 
This will involve orientifolds in the bulk, which will have a major impact on the analysis of the duality.
The analysis would be quite subtle, as anybody who has worked on orientifolds knows.

So far, we have focused on vacuum degeneracy. However, we expect more intriguing phenomena to emerge at higher energies.

Firstly, we expect black hole formation at energies of $O(N^2)$ \cite{Hawking:1982dh}. In the case of $\mathcal{N}{=}4$ super Yang-Mills theory on $S^3$, the superconformal index \cite{Romelsberger:2005eg,Kinney:2005ej} captures the entropy of supersymmetric black holes in AdS$_5$ \cite{Gutowski:2004ez,Gutowski:2004yv,Chong:2005da,Kunduri:2006ek} at high energies, thereby accounting for their microstates \cite{Cabo-Bizet:2018ehj,Choi:2018hmj,Benini:2018ywd}. By performing an analogous analysis for $S^3/\Gamma$, we may be able to elucidate the effects of a nontrivial spacetime topology on black holes, especially on the topology of the event horizon. (For recent studies on the gravity side when $\Gamma= \mathbb{Z}_k$, see \cite{Bobev:2025xan,Park:2025fon}.) One interesting point is that, in the presence of black holes, the bulk topology becomes $\bR_t \times \bR_+ \times S^3/\Gamma \times S^5$, which no longer contains the singularity.

Secondly, at sufficiently high energies such that the effects of RR 5-form flux and the curvature of $S^5$ are negligible, we expect tensionless self-dual strings of 6d $(2,0)$ theory on $S^5$ to \emph{deconfine} \cite{Choi:2018hmj,Nahmgoong:2019hko,Dorey:2022cfn}. In this regime, the supersymmetric index on $S^3/\Gamma$ would exhibit an $O(e^{k^3})$ degeneracy when $\Gamma=\bZ_k$, with an appropriate generalization for non-Abelian $\Gamma$. This should also be compared to the entropy of supersymmetric black holes in AdS$_7$ \cite{Chong:2004dy,Chow:2007ts,Bobev:2023bxl} at large $k$. 
Since the radii of AdS$_5$ and $S^5$ are equal, we expect that the deconfinement transition on $S^5$ and the Hawking-Page transition on AdS$_5/\Gamma$, discussed in the previous paragraph, to occur at a similar energy scale (or the corresponding conjugate chemical potential). It would also be illuminating to study thermodynamic phase structure on AdS$_5 /\Gamma \times S^5$, involving these two distinct phase transitions.

One promising direction towards a holographic understanding of these phenomena, including the vacuum structure, would be to recast the supersymmetric index on $S^3/\Gamma$ in light of recent developments in the giant graviton expansion \cite{Imamura:2021ytr,Gaiotto:2021xce,Murthy:2022ien}. In particular, when $\Gamma=\mathbb{Z}_k$, the matrix integral formula for the index in each holonomy sector is well-known from supersymmetric localization \cite{Benini:2011nc}. One may attempt to reorganize the index in each sector as a giant graviton expansion, and then sum over all sectors to obtain the full index. Understanding the physical meaning of such a sum in the context of the giant graviton expansion would provide valuable insights.

Finally, it would also be worthwhile to explore what we can learn in the finite-temperature setup. The authors would like to revisit some of these issues in the future. Other researchers interested in these are also welcomed to do so, of course.

\section*{Acknowledgments}
The authors thank discussions with A. Gadde, S. S. Hosseini, S. Kim, Y. Kojima, J. Lee, K.-H. Lee, S. Minwalla, K. Ohmori, T. Okazaki and J. Song.
SC and YT are both supported in part by WPI Initiative, MEXT, Japan at Kavli IPMU, the University of Tokyo.
SC is supported in part by Basic Science Research Program through the
National Research Foundation of Korea(NRF) funded by the Ministry of Education(RS-2025-02663044).
YT is supported in part by JSPS KAKENHI Grant-in-Aid (Kiban-C), No.24K06883.

\appendix

\section{Topology of $S^3/\Gamma$ and the ALE space}
\label{app:geom}

Here we collect standard facts on the topology of $S^3/\Gamma$ for finite subgroups $\Gamma\subset SU(2)$,
and the associated asymptotically locally Euclidean (ALE) space $X_\Gamma$.  
We start from the topology of $S^3/\Gamma$, then we discuss the topology of $X_\Gamma$.
These discussions will be done uniformly for all $\Gamma$,
using basic tools of algebraic topology. (See e.g.~\cite{Hatcher} for a nice introduction to this subject.)
At the end, we provide a more detailed discussion for the simplest but important case, $\Gamma=\bZ_k$,
in a much more down-to-earth manner.

\subsection{Topology of $S^3/\Gamma$}
\label{app:geom1}
Let us begin by considering $S^d/\Gamma$ in general, where
a finite group $\Gamma$ acts on $S^d$ freely.
We clearly have $H_0(S^n/\Gamma;\bZ)=\bZ$.
The theorem of Hurewicz  says that 
$H_1(S^d/\Gamma;\bZ)$ is the Abelianization of $\pi_1(S^d/\Gamma)\simeq \Gamma$,
which we denote by $\Gamma_\text{ab}$.
$H_2(S^d/\Gamma;\bZ)$ is at most torsion.
Then, the universal coefficient theorem then determines $H^{n}(S^d/\Gamma;\bZ)$
to be $\bZ$, $0$, $\hat \Gamma_\text{ab}$ for $n=0,1,2$,
where $\hat\Gamma_{ab}$ is the Pontryagin dual to $\Gamma_{ab}$,
i.e.~the group of one-dimensional representations of $\Gamma$,
with the group operation given by the tensor product. 

Note that in the main text we used the notation $I_\Gamma:=\hat\Gamma_\text{ab}$.
Let us then use $\hat I_\Gamma$ for $\Gamma_\text{ab}$.
We have so far know \begin{equation}
\begin{array}{c|cccc}
n & 0 & 1 & 2  \\
\hline
H_n(S^d/\Gamma;\bZ) & \bZ & \hat I_\Gamma & ? \\
H^n(S^d/\Gamma;\bZ) & \bZ & 0 &  I_\Gamma
\end{array}\qquad .
\end{equation}

Let us pause here to study the topological classification of $U(1)$ bundles on $S^d/\Gamma$.
In general, the topology of $U(1)$ bundles $A$ on $S^d/\Gamma$ is given in terms of 
its first Chern class $c_1(A)\in H^2(S^d/\Gamma;\bZ)$.
A flat $U(1)$ bundle  is specified by its holonomy,
given by a homomorphism $\rho: \Gamma\to U(1)$.
This is an element $\rho$ of $I_\Gamma \simeq H^1(S^d/\Gamma;U(1))$
as the Pontryagin dual of $\hat I_\Gamma \simeq H_1(S^d/\Gamma;\bZ)$.
The Bockstein $\beta: H^1(S^d/\Gamma;U(1))\to H^2(S^d/\Gamma;\bZ)$
is then also an isomorphism.
What all these mean is that the first Chern class $c_1$ of 
a flat $U(1)$ bundle $A_\rho$ specified by a homomorphism $\rho$ 
is $\rho$ itself, i.e.~$c_1(A_\rho)=\rho\in H^2(S^d/\Gamma;\bZ)$.

Let us come back to the discussion of (co)homology groups.
In our case where $d=3$, we have the Poincar\'e duality,
which provides a natural isomorphism 
$\PD: H_n(S^3/\Gamma;\bZ)\xrightarrow{\sim} H^{3-n}(S^3/\Gamma;\bZ)$.
We can then complete the table above into
\begin{equation}
\begin{array}{c|cccc}
n & 0 & 1 & 2 & 3 \\
\hline
H_n(S^3/\Gamma;\bZ) & \bZ & \hat I_\Gamma & 0 & \bZ\\
H^n(S^3/\Gamma;\bZ) & \bZ & 0 & I_\Gamma  & \bZ
\end{array}\qquad .
\label{cohoSG}
\end{equation}
Note that $\PD: H_1(S^3/\Gamma;\bZ)\xrightarrow{\sim} H^2(S^3/\Gamma;\bZ)$ provides
an isomorphism $\PD: \hat I_\Gamma \xrightarrow{\sim} I_\Gamma$.
In other words, there is a perfect pairing 
of the form \begin{equation}
I_\Gamma \times I_\Gamma \to U(1).
\label{tp}
\end{equation}
This is just the torsion (or linking) pairing on $I_\Gamma=H^2(S^3/\Gamma;\bZ)$.\footnote{%
An explicit group theoretical formula for this pairing is given as follows:
$$
\langle \rho_i, \rho_j \rangle = \frac{1}{|\Gamma|}\sum_{e\neq g\in \Gamma} 
\frac{(\rho_i(g)-1)(\rho_j(g)-1)}{(\lambda(g)-1)(\lambda(g)^{-1}-1)}\in \bR/\bZ.
$$
Here, $\lambda(g)^{\pm1}$ are two eigenvalues of $\rho_{\bC^2}(g)$,
in the standard two-dimensional representation of $\Gamma$ given 
by the inclusion into $SU(2)$.
This formula can be derived using the fact that the eta invariant $\eta(D_{\rho_i})$ 
of the Dirac operator coupled to $\rho_i$ gives a quadratic refinement of the torsion pairing,
and then using the equivariant Atiyah-Patodi-Singer theorem, see e.g.~\cite[Lemma~2.1(c)]{Gilkey}.
\label{foot:eta}
}

Now consider the standard map $S^3/\Gamma\to B\Gamma$ 
classifying the $\Gamma$ bundle $S^3\to S^3/\Gamma$.
This is a 3-equivalence, i.e.~the induced homomorphisms $\pi_k(S^3/\Gamma)\to \pi_k(B\Gamma)$ are isomorphisms for $k<3$ and surjective for $k=3$.
Therefore, the induced maps $H_k(S^3/\Gamma;\bZ) \to H_k(B\Gamma;\bZ)$ 
for $k<3$ are isomorphisms.
This in particular shows that $H_2(B\Gamma;\bZ)=0$,
from which we have $H^2(B\Gamma;U(1))=0$ via the universal coefficient theorem.
This means that the finite subgroup  $\Gamma\subset SU(2)$ does not have any
genuine projective representation, and we provided a geometric proof to it.

\subsection{Topology of the ALE space}
\label{app:geom2}

Next we study the topology of the asymptotically locally Euclidean (ALE) space,
obtained by making the singular space  $\bC^2/\Gamma$ smooth.
As always, we denote by $\sG$ the simply-connected simply-laced group corresponding to $\Gamma$ by the McKay correspondence.
It is standard that the singularity at the origin can be resolved by introducing small 2-spheres
$C_i$, $i=1,\ldots, r_\sG$, where $r_\sG=\rank \sG$. 
Let  $X_\Gamma$ denote the space $\bC^2/\Gamma$ after de-singularization. 
We still have \begin{equation}
\partial X_\Gamma=S^3/\Gamma.
\end{equation}
The small spheres $\{C_i\}$ intersect according to (the negative of) the Cartan matrix $C_{ij}$ of $\sG$: \begin{equation}
\#(C_i \cup C_j) = -C_{ij},
\end{equation} and generate the second homology group of $X_\Gamma$, $
H_2(X_\Gamma;\bZ)=\bZ^{r_\sG}
$. This allows us to identify it with the root lattice $Q_\sG$ of $\sG$, i.e.~we have \begin{equation}
H_2(X_\Gamma;\bZ)=Q_\sG.
\end{equation}
On a space $M$ with boundary, the Poincar\'e duality pairs 
$H_n(M;\bZ)$ and $H_{\dim M-n}(M,\partial M;\bZ)$,
where the latter consists of cycles which are allowed to have boundaries on $\partial M$.
In our case, this means that $H_2(X_\Gamma,\partial X_\Gamma;\bZ)$ 
is the lattice dual to $Q_\sG$, i.e.~the weight lattice $P_\sG$ of $\sG$: \begin{equation}
H_2(X_\Gamma, \partial X_\Gamma;\bZ)=P_\sG.
\end{equation}
Denoting the basis of $P_\sG$ by $\{D_i\}$, we have \begin{equation}
\#(C_i \cup D_j) =\delta_{ij}.
\end{equation}

A part of the long exact sequence of homology groups says that \begin{equation}
\begin{array}{clclclclc}
&H_2(\partial X_\Gamma;\bZ)_{=0}&\to&
H_2(X_\Gamma;\bZ)_{=Q_\sG}&\to&
H_2(X_\Gamma, \partial X_\Gamma;\bZ)_{=P_\sG}\\
\xrightarrow{(*)}&
H_1(\partial X_\Gamma;\bZ)_{=\hat I_\Gamma}&\to&
H_1( X_\Gamma;\bZ)
\end{array}
\end{equation}
is exact. 
To explain the map $(*)$,
note that $H_2(X_\Gamma;\bZ)=Q_\sG$ is a subgroup of $P_\sG$,
by regarding a 2-cycle within $X_\Gamma$ as a 2-cycle possibly with a boundary in $\partial X_\Gamma$.
Then, the elements of $P_\sG$ not in $Q_\sG$ do have boundaries in $\partial X_\Gamma$,
and determine a 1-cycle, i.e.~an element of $H_1(\partial X_\Gamma;\bZ)$.
Now, we know $P_\sG/Q_\sG=Z_\sG$,
and $|Z_\sG|=|\hat I_\Gamma|$.
This means that \begin{equation}
P_\sG/Q_\sG = Z_\sG \simeq \hat I_\Gamma,
\label{PQ1}
\end{equation}
and $H_1(X_\Gamma;\bZ)=0$.

With boundary, the Poincar\'e duality says that 
$H_n(M;\bZ)\simeq H^{\dim M-n}(M,\partial M;\bZ)$, and
$H^n(M;\bZ)\simeq H_{\dim M-n}(M,\partial M;\bZ)$.
In our case, this means that \begin{equation}
H^2(X_\Gamma;\bZ)=P_\sG,\quad
H^2(X_\Gamma,\partial X_\Gamma;\bZ)=Q_\sG.
\end{equation} The part of the long exact sequence relevant to us is \begin{equation}
H^2(X_\Gamma,\partial X_\Gamma;\bZ)_{=Q_\sG} \to 
H^2(X_\Gamma;\bZ)_{=P_\sG}
\to H^2(\partial X_\Gamma;\bZ)_{=I_\Gamma}.
\end{equation}
This sequence relates, successively, $U(1)$ bundles on $X_\Gamma$ which are trivial on $\partial X_\Gamma$,
$U(1)$ bundles on $X_\Gamma$,
and $U(1)$ bundles on the boundary $\partial X_\Gamma$.
In particular, we have \begin{equation}
P_\sG/Q_\sG =Z_\sG \simeq I_\Gamma.
\label{PQ2}
\end{equation}
Combining \eqref{PQ1} and \eqref{PQ2}, we find that the torsion pairing \eqref{tp}
is the natural $U(1)$-valued pairing on $Z_\sG$, i.e.~the $\bR$-valued pairing 
on $P_\sG\times P_\sG$ coming from the $\bZ$-valued pairing on $Q_\sG\times Q_\sG$
given by the negative $-C_{ij}$ of the Cartan matrix,
taken modulo $\bZ$.\footnote{%
It is a fun exercise to check the agreement of the pairing on $Z_\sG$
given by the negative of the inverse Cartan matrix and the pairing on $I_\Gamma$
given in footnote \ref{foot:eta}.
}

In \cite{Kronheimer}, a particularly nice basis of $H^2(X_\Gamma;\bZ)=P_\sG$ was constructed,
making the McKay correspondence manifest.
Namely, 
for each irreducible representation $\rho_i:\Gamma\to U(m_i)$,
a $U(m_i)$ instanton configuration $F_i$ on $X_\Gamma$ was constructed, 
which asymptotes to the flat bundle on $S^3/\Gamma$ specified by $\rho_i$.
Their first Chern classes $\omega_i:=c_1(F_i)$ take values in $H^2(X_\Gamma,\partial X_\Gamma;\bZ)$,
and were shown to form a dual basis to $C_j$: \begin{equation}
\int_{C_j} \omega_i=\delta_{ij}.
\end{equation} 
Now, the restriction of the first Chern class $c_1(F_i)$
on the boundary $\partial X_\Gamma=S^3/\Gamma$ is by definition
the first Chern class of the determinant bundle $\det \rho_i$ of the flat bundle $\rho_i$.
This gives the element $\det\rho_i \in I_\Gamma \simeq H^2(S^3/\Gamma;\bZ)$,
explicitly giving the map $P_\sG = H^2(X_\Gamma;\bZ)\to H^2(\partial S^3/\Gamma;\bZ)=I_\Gamma$.

\subsection{A concrete study of the case $\Gamma=\bZ_k$}

Finally, let us consider $S^3/\bZ_k$ and $X_{\bZ_k}$ in a more concrete way.
We parameterize $\bC^2$ by two complex variables $(u,v)$,
and let $\bZ_k$ to act via \begin{equation}
(u,v) \mapsto (\omega u,\omega^{-1} v),\qquad \omega=e^{2\pi i/k}.
\label{uv}
\end{equation}
This also determines an action of $\bZ_k$ on the unit sphere $S^3\subset \bC^2$.

In the case of $S^3/\bZ_k$, the computation of the (co)homology groups
is most easily and explicitly done by realizing the space as a cell complex.
It turns out that we only need one cell in each dimension, and the boundary maps can also be explicitly determined, see e.g.~Example 2.43 of \cite{Hatcher}.
The complex of the chain groups $C_n$ to compute $H_n(S^3/\bZ_k;A)$ for an Abelian group $A$ is then 
\begin{equation}
C_0 \xrightarrow{0} 
C_1 \xrightarrow{3\times }
C_2 \xrightarrow{0}
C_3
\end{equation} and the complex of the cochain groups $C^n$ to compute $H^n(S^3/\bZ_k;A)$ is similarly
given by \begin{equation}
C^0 \xleftarrow{0} 
C^1 \xleftarrow{3\times }
C^2 \xleftarrow{0}
C^3,
\end{equation} where all $C^i$ and $C_i$ are simply $A$.
Taking $A=\bZ$, we reproduce \eqref{cohoSG} in this manner, where $I_\Gamma=\hat I_\Gamma=\bZ_k$.

Let us now move on to the study of $\bC^2/\bZ_k$ and its smooth version $X_{\bZ_k}$.
The explicit metric is known, but for our purposes we only need the topology,
which can be usefully studied in the following manner, using complex coordinates.
All of the discussions below are standard in mathematics; 
it can also be found in many places in string theory literature, e.g.~\cite{Hori:1997zj},
where other cases of $\Gamma$ are also treated in a similar manner. 

Instead of the coordinates $(u,v)$ with the identification \eqref{uv},
let us use the $\bZ_k$-invariant combinations \begin{equation}
x:=u^k,\quad y:=v^k, \quad z:=uv,
\end{equation} satisfying \begin{equation}
xy = z^k.
\end{equation} This realizes $\bC^2/\bZ_k$ as a hypersurface in $\bC^3$ parameterized by $(x,y,z)$,
with a singularity  at the origin.

\begin{figure}
\centering
\includegraphics[width=.6\textwidth]{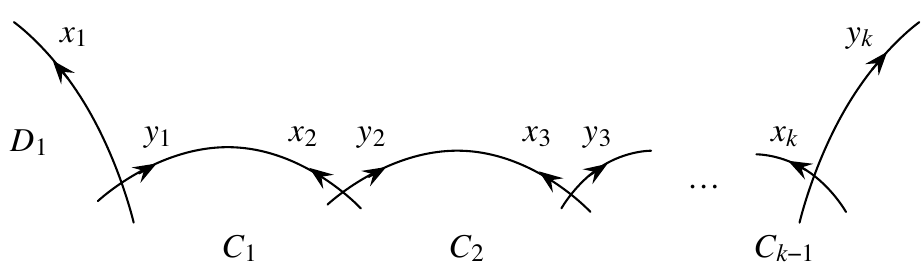}
\caption{The space $X_{\bZ_k}$, consisting of $k$ patches $U_i \simeq\bC^2$ with coordinates 
$(x_i,y_i)$, $(i=1,\ldots, k)$, where we draw complex planes as real lines.
We also indicated $k$ 2-spheres $C_{1,\ldots, k}$,
and one non-compact 2-cycle $D_1$.
\label{fig:XZ}}
\end{figure}

We blow up the origin $k-1$ times, producing the space $X_\Gamma$.
It has $k$ open patches  $U_i\simeq \bC^2$ for $i=1,\ldots, k$,
with coordinates $(x_1,y_1)$, $(x_2,y_2)$, \ldots, $(x_k,y_k)$.
We paste the patches by demanding \begin{equation}
y_1 x_2 = y_2 x_3 = \cdots = y_{k-1} x_k =1 
\label{yx}
\end{equation} and \begin{equation}
x_1 y_1 = x_2 y_2 = \cdots = x_k y_k.
\label{xy}
\end{equation}
We define the projection \begin{equation}
\pi: X_{\bZ_k} \to \bC^2/\bZ_k
\end{equation}
by \begin{equation}
x=x_1, \quad y=y_k, \quad z = x_1y_1= \cdots = x_k y_k.
\end{equation} It is straightforward to check that $xy=z^k$ is indeed satisfied,
by rewriting $x_1 y_1 x_2 y_2 \cdots x_k y_k$ in two ways,
one using \eqref{yx}, another using \eqref{xy}. 
See Fig.~\ref{fig:XZ} for a schematic depiction of the space.

Let $C_i$ for $i=1,\ldots,k-1$ be defined as \begin{equation}
C_i = \{  (x_i,y_i)\in U_i \mid x_i =0  \} \cup
\{  (x_{i+1},y_{i+1})\in U_{i+1} \mid y_{i+1} =0  \}.
\end{equation} 
It is easy to see that $C_i$ is a $\CP^1$ parameterized by $y_i=1/x_{i+1}$,
and is in the inverse image of the origin $(x,y,z)=(0,0,0)$  of the projection $\pi$.
Therefore, the space $X_{\bZ_k}$ is indeed obtained by $\bC^2/\bZ_k$
by inserting $(k-1)$ two-spheres at the origin.
We can directly see that 
$C_i$ and $C_j$ for $i\neq j$ intersect according to the Dynkin diagram of $A_{k-1}=SU(k)$;
with a more effort, we can also check that the self-intersection number of $C_i$ with itself is $-2$.
This shows that $\#(C_i\cup C_j)$ is given by the negative of the Cartan matrix of $SU(k)$,
and that the set $\{C_i\}$ generates the root lattice $Q_{SU(k)}$.

An example of a 2-cycle in the dual lattice is the subspace \begin{equation}
D_1 := \{ (x_1,y_1)\in U_1 \mid y_1=0 \}.
\end{equation} This intersects only with $C_1$, and so corresponds to a
fundamental weight vector in $P_{SU(k)}$.
Its projection to $\bC^2/\bZ_k$ is given by $y=0$, or $v=0$ in the original variable. 

For explicitness, let us consider the boundary $S^3/\Gamma$
to be at a finite distance rather than at asymptotic infinity, by saying that $S^3$ has the radius $R\gg 1$.
Then, the intersection of $D_1$ with the boundary is the boundary circle \begin{equation}
\{  (u,v) \in S^3 \mid  |u| =R \} ,\quad\text{with the identification}\quad  u \sim e^{2\pi i/k} u  ,
\end{equation}  which is a generator of $\pi_1(S^3/\bZ_k )=H_1(S^3/\bZ_k;\bZ)=\bZ_k$.

Then, the boundary of $kD_1$ should be homologically trivial, 
and therefore $kD_1$ should be a linear combination of $C_i$.
From the condition that $\#(kD_1 \cup C_i)=k$ for $i=1$ and zero otherwise, we should have \begin{equation}
kD_1 \sim - [ (k-1) C_1 + (k-2) C_2 + \cdots + C_{k-1}]
\label{HW}
\end{equation} as a homology cycle. 
Let us show this explicitly. 

For this purpose, consider the surface $S_c$ defined by the equation $y=c$ first for $c\neq 0$, and let $c\to 0$.
When $c\neq 0$, the surface $S_c$ avoids the singularity, and its topology 
can be studied in $\bC^2/\bZ_k$.
Again for explicitness, consider only the region where $|u|^2+|v|^2\le R^2$.
Recalling that $y=v^k$, the equation $y=c$ determines the region \begin{equation}
\{|u|^2 \le R^2-|c|^{2/k}  \}  \times \{v \mid v=e^{2\pi \ell/k}c^{1/k}\}
\end{equation} under the identification $(u,v)\sim (e^{2\pi i/k} u,e^{-2\pi i/k v})$.
This identification moves all the allowed choices of $v$.
So the surface $S_c$ is simply a disk of radius $\sqrt{R^2-|c|^{2/k}}$
as long as $|c|>0$, which smoothly disappears as we send $|c|\to R$.

Instead, if we take $|c|\to 0$, $v$ becomes zero, and 
the $\bZ_k$ identification starts to act on the disk $\{|u|=R \} $ of radius $R$
by $u\sim e^{2\pi i/k} u$.
Therefore the $|c|\to 0$ limit of $S_c$ gives $k$ times the pizza slice $B^2/\bZ_k$,
in the singular space $\bC^2/\bZ_k$.
To see what happens in $X_{\bZ_k}$, we simply rewrite the equation $y=c$ in each of the coordinate patches $U_i$.
We find \begin{equation}
c=y= y_k = x_{k-1} y_{k-1}^2 = x_{k-2}^2 y_{k-2}^3 = \cdots = x_1^{k-1} y_1^k.
\end{equation}
This means that the limit $c\to 0$ of $S_c$ in the patch $U_i$
wraps the locus $\{x_i=0\}$ $(k-i)$ times and the locus $\{y_i=0\}$ $(k-i+1)$ times.
In other words,
the limit $c\to 0$ of $S_c$ wraps
$D_1$ $k$ times,
$C_1$ $(k-1)$ times,
$C_2$ $(k-2)$ times,
\ldots,
$C_{k-1}$ one time.
This way, we succeeded in realizing a continuous deformation between
$k D_1 + (k-1) C_1 + \cdots + C_{k-1}$
and an empty cycle in terms of $S_c$.
That is, we have demonstrated the equality \eqref{HW}.

\section{Details of the McKay correspondence on $I_\Gamma$ and $Z_\sG$ }
\label{sec:checks}
In Sec.~\ref{sec:1-form-matching}, the agreement of the actions of 1-form symmetries
on the side of \Nequals4 super Yang-Mills on $S^3/\Gamma$
and on the side of pure Chern-Simons theory with gauge group $\sG$ on $T^2$
was reduced to the two mathematical statements, \eqref{E-nontrivial} and \eqref{M-nontrivial},
relating the properties of representations of $\Gamma$
and weight vectors of $\sG$.
These statements should not be new, but the authors are not aware of a place
where they are explicitly mentioned.
Here we provide a direct check of the statements.

\subsection{Statements to be established}
Let us first recall the notations, so that this appendix can be read
without flipping the pages of this paper to locate the definitions.
Let $\Gamma$  be a finite subgroup of $SU(2)$,
$\rho_{i=1,\ldots, \ell_\Gamma}$ be the irreducible representations 
and $I_\Gamma$ be the set (and therefore the group) of one-dimensional irreducible representations.
We let $\rho_{i=1}$ be the trivial representation. 
The geometry of $S^3/\Gamma$ provides a natural perfect pairing 
\begin{equation}
I_\Gamma\times I_\Gamma\to U(1).\label{Ipair}
\end{equation}

Let $\sG$ be the corresponding simple simply-connected Lie group of ADE type,
$P_\sG$, $Q_\sG$ be the weight and the root lattice, and 
$Z_\sG=P_\sG/Q_\sG$ be the center of $\sG$.
There is a natural perfect pairing \begin{equation}
Z_\sG \times Z_\sG\to U(1),\label{Zpair}
\end{equation}
which descends from the pairing on $Q_\sG$, which is the negative
of the Cartan matrix in our convention.
As we recalled in Appendix~\ref{app:geom},
the geometry of $S^3/\Gamma$ and the ALE space
provides a natural isomorphism $Z_\sG \simeq I_\Gamma$
between Abelian groups with a perfect pairing.\footnote{%
Note that the agreement of the pairing contains more information than the isomorphism
of the groups.
For example, for both $\sG=SU(3)$ and $E_6$,
$Z_\sG=\bZ_3$,
but the pairing of the generator with itself is $+1/3$ for the former while $-1/3$ for the latter.
Also, for both $\sG=SO(8k)$ and $SO(8k+4)$, $Z_\sG=\bZ_2\times \bZ_2$,
but the pairing is given by $\langle x,y\rangle=1/2$ for the former
while $\langle x,x\rangle=\langle y,y\rangle=1/2$ for the latter,
where $x$, $y$ are the generators of two factors of $\bZ_2\times \bZ_2$.
}

Let $\pi_{i=1,\ldots, r_\sG+1}$ be the affine weight vectors,
corresponding to the nodes of the extended Dynkin diagram.
Let $m_i$ be the comarks. 
Then we have $m_i=\dim \rho_i$ under the McKay correspondence.
This means that we can naturally identify the subset \begin{equation}
 I_\sG:= \{\pi_i \mid m_i=1\} \subset \{\pi_i\} \subset P_\sG
\end{equation}
with $I_\Gamma$. 
We can now consider the map
\begin{equation}
I_\sG\ni  \pi_i  \mapsto [\pi_i] \in P_\sG/Q_\sG \simeq Z_\sG.
\end{equation}
This turns out to be an isomorphism, 
i.e.~the subset $I_\sG$ provides 
a particularly nice set of representatives of the equivalence classes $Z_\sG \simeq P_\sG /Q_\sG$.
With this, we can define a map $\zeta$, \begin{equation}
\{\pi_i\} \ni \pi_i \mapsto \pi_{\zeta(i)} \in I_\sG, \label{w}
\end{equation} by the condition \begin{equation}
[\pi_i] = [\pi_{\zeta(i)}] \in Z_\sG\simeq P_\sG/Q_\sG.
\end{equation}

Finally, let $O_\sG$ be the outer automorphism group of the affine Lie algebra of type $\sG$.
An element $o\in O_\sG$ acts on the set $\{\pi_i\}$ by permutation, $\pi_i \mapsto \pi_{o(i)}$,
preserving the comarks, $m_i$.
We then have the isomorphism \begin{equation}
O_\sG \ni o \mapsto \pi_{o(1)}\in I_\sG \simeq Z_\sG.
\end{equation} 

We are going to check two  nontrivial agreements concerning $I_\Gamma$ and $Z_\sG$. 
The first is the relation \eqref{E-nontrivial}, which is \begin{equation}
\rho_{o(1)} \otimes \rho_i = \rho_{o(i)}, \qquad \text{for all}\ o\in O_\sG.
\label{to-be-checked-E}
\end{equation}  
The non-triviality here is that the operations on the left and on the right are determined 
on the side of $\Gamma$ and on the side of $\sG$, respectively.

The second is the relation \eqref{M-nontrivial}, which is \begin{equation}
\omega_z(\det\rho_i)=z(\pi_i), \label{oz}
\end{equation} where $z\in Z_\sG$ and $\omega_z \in \hat I_\Gamma$ 
is the element corresponding to $z$ under the isomorphism 
$Z_\sG \simeq I_\Gamma \simeq \hat I_\Gamma$,
and the second isomorphism is in terms of the pairing $I_\Gamma\times I_\Gamma \to U(1)$.
Let us rewrite this relation in a way which is easier to be checked. 
On the left hand side, $\det\rho_i\in I_\Gamma$.
On the right hand side, $z(\pi_i)$ only depends on the class $[\pi_i] \in Z_\sG$.
We discussed above in \eqref{w} that there is a unique element $\pi_{\zeta(i)}\in I_\sG$ such that 
$[\pi_i]= [\pi_{\zeta(i)}]$.
The relation \eqref{oz} then simply means that \begin{equation}
\det\rho_i = \rho_{\zeta(i)}.
\label{to-be-checked-M}
\end{equation}
Again, the non-triviality is that the operations on the left and on the right are computed 
on the side of $\Gamma$ and on the side of $\sG$, respectively.
This relation actually follows from Kronheimer's construction, as recalled at the end of Appendix~\ref{app:geom2}.
We will check the relation directly below, nonetheless.
Note that, as $\zeta$ and $\det$ are clearly  identity operations for $\pi_i\in I_\sG$
and $\rho_i\in I_\Gamma$, 
we only have to check it for $\rho_i$ or $\pi_i$ outside of $I_\sG$,
i.e.~for those $m_i=\det\rho_i >1$.

\subsection{Individual cases}

Let us now study the individual cases, one by one.
The reader should consult Table~\ref{tab:McKay} and Table~\ref{tab:diagram}
for various basic data of the McKay correspondence.

\paragraph{Type A.}
We first consider the case $\Gamma=\bZ_k$ and $\sG=SU(k)$.
The extended Dynkin diagram is a circle of $k$ nodes.
Let us use $\tilde\rho_j$, ($j=1,\ldots,k-1$), to denote the one-dimensional representation
where the generator of $\bZ_k$ acts by $e^{2\pi i j/k}$.
Then the edges connect $\tilde\rho_j$ and $\tilde\rho_{j+1}$, 
where the subscripts are identified modulo $k$.
(In the main text, the trivial representation was defined to sit at $\rho_1$.
So we have $\rho_i = \tilde \rho_{i-1}$.)

To check \eqref{to-be-checked-E}, it is enough to take $o$ to be the generator of $O_\sG=\bZ_k$.
This outer automorphism action rotates the circle by one unit,
and clearly agrees with tensoring with $\rho_{o(1)}=\tilde \rho_1$.
To check \eqref{to-be-checked-M}, we simply note that $\det$ is a trivial operation
for one-dimensional representations,
and that $\zeta$ is also the identity map.

\paragraph{Type D.}
Let us next consider the case of $\Gamma=\hat D_{m}$ and $\sG=D_{m+2}$.
The group $\hat D_{m}$ is generated by two elements $a$ and $b$, satisfying \begin{equation}
a^2=b^{2m}=(ab)^2.
\end{equation}
There are $m-1$ two-dimensional representations $\rho_{2_k}$ for $k=1,\ldots, m-1$, given by 
\begin{equation}
\rho_{2_k}(a)=\begin{pmatrix}0&i^k\\i^k&0\end{pmatrix},\qquad
\rho_{2_k}(b)=\begin{pmatrix}\alpha^k&0\\0&\alpha^{-k}\end{pmatrix}\ ,
\end{equation}
where $\alpha=e^{\pi i/m}$.
We can similarly define $\rho_{2_0}$ and $\rho_{2_m}$, but they decompose into two one-dimensional representations:
\begin{equation}
	\rho_{2_0}=\rho_1\oplus \rho_{1'}\ ,\quad
	\rho_{2_m}=\rho_{1''}\oplus \rho_{1'''}\ .
\end{equation}
Explicitly, we have 
\begin{equation}
	\begin{array}{c|cccc}
		& 1 & 1' & 1'' & 1'''\\
		\hline
		a & 1 & -1 & i^m & -i^m\\
		b & 1 & 1 & -1 & -1
	\end{array}\ .
\end{equation}
These are the irreducible representations of $\hat\cD_m$.
They form the extended Dynkin diagram of type $D_{m+2}$: 
\begin{equation}
 \bnode{}{1}-\node{\ver{}{1'}}{2_1}-\node{}{2_2}-\cdots-\node{\ver{}{1''}}{2_{m-1}}-\node{}{1'''} \ .
\end{equation}
Below, we use the names given above as the subscripts for $\rho_i$ and $\pi_i$,
instead of $i=1,\ldots, m+2$.

Let us now check the statements \eqref{E-nontrivial} and \eqref{M-nontrivial}.
As for the statement \eqref{E-nontrivial}, $I_\sG$ is generated by  $1'$ and $1''$, 
so we have to check their actions, on the side of $\Gamma$ and on the side of $\sG$.
On the side of $\Gamma$, tensoring by $1'$ fixes all $2_x$'s,
and exchanges the four 1-dimensional representations as 
$1\leftrightarrow 1'$ and $1''\leftrightarrow 1'''$.
Tensoring by $1''$ exchanges $2_{k}\leftrightarrow 2_{m-k}$, and 
send $1''\to 1$ when $m$ is even while send $1''\to 1'$ when $m$ is odd.
On the side of $\sG$, the actions of the outer automorphism group
are given e.g.~in \cite[Fig.~14.1, Table 14.1]{DiFrancesco:1997nk}.
We can then easily check the agreement of the actions on the side of $\Gamma$
and the actions on the side of $\sG$.

Let us turn to the statement \eqref{M-nontrivial}.
On the side of $\Gamma$, $\det\rho_i$ can be computed explicitly from the data given above.
$\det$ is clearly an identity on one-dimensional representations,
and $\det\rho_{2_k}=\rho_1$ or $\rho_{1'}$ depending on whether $k$ is odd or even.
On the side of $\sG$, we have \begin{equation}
\begin{aligned}
\pi_{1'}&=(1,0,0,\ldots,0,0,0),\\
\pi_{2_1}&=(1,1,0,\ldots,0,0,0),\\
\pi_{2_2}&=(1,1,1,\ldots,0,0,0),\\
& \vdots \\
\pi_{2_{m-1}}&=(1,1,1,\ldots,1,0,0),\\
\pi_{1''}&=(\tfrac12,\tfrac12,\tfrac12,\ldots,\tfrac12,\tfrac12,+\tfrac12),\\
\pi_{1'''}&=(\tfrac12,\tfrac12,\tfrac12,\ldots,\tfrac12,\tfrac12,-\tfrac12)\ .
\end{aligned}
\end{equation}
As we already mentioned, $\zeta(i)=i$ for $i\in I_\sG$, 
i.e.~for $i=1, 1', 1''$ and $1'''$.
We can also check that $\zeta(2_k)=1$ or $1'$, depending on whether $k$ is odd or even.
Thus we see that the actions of $\det$ and $\zeta$ agree.

\paragraph{Type E.}
Let us finally study the exceptional cases.
We start with $E_6$.
Name the nodes as follows:
\begin{equation}
\bnode{}{1}-\node{}{2}-\node{\overset{\ver{}{1'}}{\ver{}{2'}}}{3}-\node{}{2''}-\node{}{1''}\ .
\end{equation}
To see the agreement of the outer automorphism action
and the tensoring by the one-dimensional representations,
it suffices to recall how McKay built the extended Dynkin diagram 
on the side of $\Gamma$. Namely, two nodes $\rho_i$ and $\rho_j$
are connected only when $\rho_i\otimes \rho_2$ contained $\rho_j$.
Then $\rho_{1'}\otimes \rho_2=\rho_{2'}$, and $\rho_{1''}\otimes \rho_2 =\rho_{2''}$.
Combined with $\rho_{1'}\otimes \rho_{1'}=\rho_{1''}$,
we see that tensoring by $\rho_{1'}$ generates the $\bZ_3$ graph automorphism,
which is the outer automorphism action.

To see the agreement of $\det$ and $\zeta$,
we first study $\det$.  Recall again $\rho_i \otimes \rho_2 = \bigoplus_{j} \rho_j$,
where $j$ runs over nodes connected to $i$. Taking the determinant on both sides,
we see $(\det\rho_i )^2= \bigotimes \det\rho_j$. 
This gives the image of $\det$ as \begin{equation}
\bnode{}{1}-\node{}{1}-\node{\overset{\ver{}{1'}}{\ver{}{1''}}}{1}-\node{}{1'}-\node{}{1''}\ .
\label{B18}
\end{equation}
To compute $\zeta$, we simply compute $\langle \pi_i,\pi_{1'}\rangle$
by inverting the Cartan matrix. We find that it is given by \begin{equation}
\bnode{}{0}-\node{}{1}-\node{\overset{\ver{}{4/3}}{\ver{}{5/3}}}{2}-\node{}{4/3}-\node{}{2/3}\ .
\end{equation}
Taking modulo 1, we see that it is compatible with \eqref{B18}.

For $E_7$,
we name the nodes as follows:
\begin{equation}
\bnode{}{1}-\node{}{2}-\node{}{3}-\node{\ver{}{2''}}{4}-\node{}{3'}-\node{}{2'}-\node{}{1'}\ .
\end{equation}
That tensoring by $\rho_{1'}$ generates the $\bZ_2$ graph automorphism proceeds
exactly as in the case for $E_6$.
The determinant can similarly be computed, and gives \begin{equation}
\bnode{}{1}-\node{}{1}-\node{}{1}-\node{\ver{}{1'}}{1}-\node{}{1'}-\node{}{1}-\node{}{1'}\ .
\label{B21}
\end{equation}
To compute $\zeta$, we compute $\langle \pi_i,\pi_{1'}\rangle$ by inverting the
Cartan matrix. We find that it is given by \begin{equation}
\bnode{}{0}-\node{}{1}-\node{}{2}-\node{\ver{}{3/2}}{3}-\node{}{5/2}-\node{}{2}-\node{}{3/2}\ .
\end{equation}
Taking modulo 1, we again see that it is compatible with \eqref{B21}.

Finally, there is nothing to be checked for $E_8$, since $I_\Gamma$ and $Z_\sG$ are both trivial.

 \bibliographystyle{ytamsalpha}
 \def\arxivfont{\rm}
 \baselineskip=.95\baselineskip
\bibliography{refs}

\end{document}